\documentclass{amsart} 
\usepackage[latin1]{inputenc}
\usepackage[english]{babel} 
\usepackage{ifpdf} 
\ifpdf
\usepackage[pdftex]{graphicx} 
\usepackage{epstopdf} 
\else
\usepackage[dvips]{graphicx}  
\fi 
\usepackage{caption} 
\usepackage{multirow}
\usepackage[pdftitle={Uncertainties in X-ray 3D reconstructions},
  pdfauthor={Engblom/Nettelblad/Liu},
  pdffitwindow=true,
  breaklinks=true,
  colorlinks=true,
  urlcolor=blue,
  linkcolor=red,
  citecolor=red,
  anchorcolor=red]{hyperref}
\usepackage{amssymb}
\usepackage{amsmath}
\usepackage{amsthm}

\usepackage{algorithmic}
\usepackage[ruled,norelsize]{algorithm2e}

\usepackage{subfig} 
\usepackage{listings}
\usepackage{mathrsfs}	
\usepackage[numbers,sort]{natbib} 
\usepackage{amsmath,mathrsfs}
\usepackage[inline]{enumitem}

\newcommand{\Mpix}{M_{\mbox{{\tiny pix}}}}
\newcommand{\Mrot}{M_{\mbox{{\tiny rot}}}} 
\newcommand{\Mdata}{M_{\mbox{{\tiny data}}}}
\newcommand{\Mgrid}{M_{\mbox{{\tiny grid}}}}

\newcommand{\Prob}{\mathbf{P}} 

\newcommand{\Wgrid}{\mathbb{W}}

\newcommand{\rvec}{\mathbf{r}}
\newcommand{\derivate}{\nabla}

\newcommand{\wavenumber}{\mathit{k}}
\newcommand{\refInd}{\mathit{n}}

\newcommand{\ppd}[1]{#1_{\perp}}
\newcommand{\Fourier} {\mathscr{F}}

\newcommand{\angstrom}{\mbox{\AA ngstr\"om}}
\newcommand{\po}{\mbox{Po}}

\numberwithin{equation}{section} 
\numberwithin{figure}{section}
\numberwithin{table}{section}

\begin{document}

\title[Uncertainties in X-ray 3D reconstructions]{Assessing
  Uncertainties in X-ray Single-particle Three-dimensional
  reconstructions}
	
\author[S. Engblom]{Stefan Engblom} 

\author[C. Nettelblad]{Carl Nettelblad} 
	
\author[J. Liu]{Jing Liu}
	
\address[S. Engblom \and J.~Liu]{Division of
  Scientific Computing, Department of Information Technology, Uppsala
  university, SE-751 05 Uppsala, Sweden.}
	

\email{stefane, jing.liu@it.uu.se}

\address[C.~Nettelblad]{Science for Life Laboratory, Division of
	Scientific Computing, Department of Information Technology, Uppsala
	university, SE-751 05 Uppsala, Sweden.}

\email{carl.nettelblad@it.uu.se}
	
\address[J.~Liu]{Laboratory of Molecular Biophysics, Department of
  Cell and Molecular Biology, Uppsala university, SE-751 24 Uppsala,
  Sweden.}
	
\email{jing.liu@icm.uu.se}
	
\thanks{Corresponding author: S. Engblom, telephone +46-18-471 27 54,
  fax +46-18-51 19 25.}

\date{\today}
	
	
\selectlanguage{english}
	






\begin{abstract}
  Modern technology for producing extremely bright and coherent X-ray
  laser pulses provides the possibility to acquire a large number of
  diffraction patterns from individual biological nanoparticles,
  including proteins, viruses, and DNA. These two-dimensional
  diffraction patterns can be practically reconstructed and retrieved
  down to a resolution of a few \angstrom. In principle, a
  sufficiently large collection of diffraction patterns will contain
  the required information for a full three-dimensional reconstruction
  of the biomolecule. The computational methodology for this
  reconstruction task is still under development and highly resolved
  reconstructions have not yet been produced.

  We analyze the Expansion-Maximization-Compression scheme, the
  current state of the art approach for this very challenging application, by
  isolating different sources of uncertainty.
  Through numerical experiments on synthetic data we evaluate their
  respective impact. We reach conclusions of relevance
  for handling actual experimental data, as well as pointing out
  certain improvements to the underlying estimation algorithm.

  We also introduce a practically applicable computational methodology
  in the form of bootstrap procedures for assessing reconstruction
  uncertainty in the real data case. We evaluate the
  sharpness of this approach and argue that this type of procedure
  will be critical in the near future when handling the increasing
  amount of data.
\end{abstract}

\maketitle


\section{Introduction}
\label{sec:intro}

Determing the structure of a very small biological object, such as a
protein or a virus, is both fascinating and hard. The most classical
and common way to determine the atomic and molecular structures of
small biological objects is to crystallize them and use X-rays to
investigate the resulting macroscopic crystals.  This method, X-ray
crystallography, has succeeded in determining more than 97,200
structures \cite{pdbsta}. With X-ray crystallography, high-quality
structures can be obtained from crystals whose atoms are formed in a
near perfect periodic arrangement. However, due to conformational
flexibility not all biological samples can form crystals.

Modern X-ray Free Electron Laser (X-FEL) technology potentially
provides the ability to determine biological structure without
crystals. X-FEL pulses are intense and short enough to create an
observable diffraction signal from one single particle, outrunning the
radiation damage. Digital detectors are used to capture the diffracted
wave, depicting the sample before it explodes and turns into a
plasma. This approach is called ``diffract and destroy'' \cite{dad},
and has caught considerable attention in structural biology
\cite{44499,394011,44491,306934,515159,3d_mimi}.

For a Flash X-ray single particle diffraction Imaging (FXI)
experiment, a stream of particles are injected into the X-ray beam,
and hit by the extremely intense X-ray pulses, producing diffraction
patterns showing the illuminated objects. The energy from the X-ray
pulse destroys the sample, so it is impossible to collect successive
exposures of the same particle. However, since many biological
particles exist in identical copies at the resolution scales of
relevance, the diffraction patterns can be treated approximately as
differently oriented exposures of the same particle. The particle
rotations can be recovered \cite{EMC,dm,gtm} by maximizing the fit
among all diffraction patterns. Hence, a 3D intensity can be assembled
as an average of these oriented patterns.

In 2011, a 2D reconstruction of a mimivirus \cite{mimivirus_Xrays} was
reported, one of the largest known viruses at a diameter of roughly
500 nm. The reconstruction was based on individual FXI diffraction
patterns, with 32-nm full-period resolution. Later, a corresponding 3D
reconstruction was also presented \cite{3d_mimi}, whose resolution was
markedly inferior to the one achieved in 2D from individual
patterns. The authors of the 3D reconstruction suggested that a
higher-resolution 3D reconstruction could be obtained by adding more
diffraction patterns from homogeneous samples. This would clearly
require a large and high quality dataset along with a comprehensive
understanding of the uncertainty propagation in the reconstruction
procedure.

In this paper, we attempt to analyze sources and propagation of
uncertainties in the Expansion-Maximization-Compression (EMC)
algorithm \cite{EMC,EMC2}, the best-in-practice 3D reconstruction
method, in order to be able to estimate the 3D reconstruction
resolution.

An overview of the computational methodology using FXI images is found
in \S \ref{sec:theory}. We discuss the sources of errors and introduce
two bootstrap schemes for estimating the reconstruction uncertainty in
\S \ref{sec:boot}. Numerical experiments to assess the impact of the
various sources of uncertainty are presented in \S \ref{sec:exp},
where we also evaluate the sharpness of the bootstrap estimators and
the robustness of the overall reconstruction procedure. A concluding
discussion is found in \S \ref{sec:conclusions}.


\section{Imaging via FXI}
\label{sec:theory}

It is well known that one can use the Fourier transform to approximate
diffracted waves in the far-field \cite{f1}. In this section, we
review the relationship between the solutions to the wave equation as
represented via Fourier transforms and the captured FXI diffraction
images. We also review Maximum Likelihood-based image processing
techniques, and the best-in-practice 3D FXI reconstruction algorithm.

\subsection{Scattering theory}
\label{subsec:scattering}
X-FEL pulses can produce diffraction patterns of single biomolecules. This 
diffraction process, a wave propagation in free space, is described by the 
Helmholtz wave equation,
\begin{align}
 \derivate^2 \Psi + \wavenumber^2 \refInd^2\Psi &= 0,
 \label{eq:helm}
\end{align}
where the wave number is $\wavenumber$, and the refractive index is $\refInd$. 

We can solve the wave equation on an Ewald sphere, which is perpendicular to 
the X-ray beam direction, by assuming that:
\begin{enumerate*}[label= \roman*)]
\item the polarization of X-FEL pulses can be ignored;
\item the objects in the X-ray beam are small, so that photons
  diffract only once inside the object (i.e.~only the first order Born
  approximation $\Psi_1$ is required);
\item small-angle scattering is assumed, i.e.~that the object-detector
  distance is much longer than the wavelength;
\item the X-FEL sources generate plane, coherent, and homogeneous
  waves.
\end{enumerate*}

The scattered wave on the Ewald sphere can then be written as follows:
\begin{align}
\bar{\Psi} \approx \sqrt{I_0 \Omega_p} \frac{2 \pi}{\lambda^2} \Fourier_{2} \{\delta\ppd{\refInd}(\ppd{\rvec})\} \Delta x^3 \propto \Fourier_{2} \{\delta\ppd{\refInd}(\ppd{\rvec})\},
\label{eq:wave}
\end{align}
 where $\Fourier_{2}$ is a 2D Fourier transformation, $\Delta x$ is
 the sampling distance, $\delta\ppd{\refInd}(\ppd{\rvec})$ is the
 refractive component of the refractive index for the Ewald sphere,
 $I_0$ is the X-ray pulse intensity at the object-beam interaction
 point, and $\lambda$ is the wave length. Further, $\Omega_p$ equals
 to $P^2 / D^2$, where $P$ is the physical pitch of a single detector
 pixel, and $D$ is the object-detector distance.
 
The noiseless diffraction pattern detected by the detector is the square of 
this scattered wave, i.e.~
\begin{align}
I = |\bar{\Psi}|^2 \propto |\Fourier_{2} \{\delta\ppd{\refInd}(\ppd{\rvec})\}|^2.
\label{eq:int}
\end{align}

We denote a collection of noiseless diffraction patterns by $K^* =
(K^*_k)_{k=1}^{\Mdata}$, where each frame $K^*_k$ is obtained from
\eqref{eq:int} by specifying effectively a rotation of the
object. Since the X-ray pulse intensity at the object-beam interaction
point $I_0$ will vary in practice, we denote this variation by $\phi$
- the (photon) fluence, such that diffraction pattern with varying
fluence is obtained by scaling $\phi K^*_k$. Moreover, since digital
detectors are pixelized, we also discretize each diffraction pattern
and write $K^*_k = (K^*_{ik})_{i=1}^{\Mpix}$, where $\Mpix$ is the
number of pixels.
 
\subsection{Maximum-Likelihood-based Imaging for FXI} 

In \eqref{eq:wave}, the refractive component of the refractive index
for the Ewald sphere $\delta\ppd{\refInd}(\ppd{\rvec})$ is dependent
on the rotation of particle. This is directly observable from FXI
experiments. Several methods \cite{EMC,gtm,dm} can be used to estimate
the unknown particle rotations from FXI diffraction patterns, but the
most successful approach so far is the EMC algorithm
\cite{3d_mimi,EMC2,EMC}. Besides calculating maximum-likelihood (ML)
estimates, the EMC algorithm interpolates between 2D diffraction
patterns and a 3D model.

The EMC algorithm consists of 4 steps per iteration: i) the expansion
step (\textbf{e step}) slices the 3D model through the model center
according to the sampled rotation, i.e.~expands the 3D model into a
set of 2D slices; ii) the expectation step (\textbf{E step}) estimates
the probability of each pattern to be in any given rotation; iii) the
maximization step (\textbf{M step}) updates the 2D slices and their
fluences using the estimated rotational probability; iv) the
compression step (\textbf{C step}) inserts the updated 2D slices back
into the 3D model.

We first introduce the e and the C step, which interpolates between a
3D model and 2D slices. Let $\Wgrid = \{\Wgrid_l\}_{l=1}^{\Mgrid}$ be
a 3D discrete model, an estimation of the 3D Fourier intensity of a
biomolecule, where $ \Mgrid = \Mpix^{3/2}$. The rotational space $R$
is discretized by $(R_j)_{j=1} ^{\Mrot}$, and the corresponding prior
weight for rotation $R_j$ is $w_j$, normalized such that $\sum_j w_j =
1$. Similarly, the intensity space is discretized by a set of pixels
$(q_i)_{i=1}^{\Mpix}$, such that the unknown 2D Fourier intensity at
position $R_jq_i$ can be denoted by $W_{ij}$ in this coordinate
system. We define interpolation weights $f$ and interpolation
abscissas $(p_l)_{l=1}^{\Mgrid}$ such that for $g$ some smooth
function,
\begin{align}
g(q) \approx \sum_{l = 1}^{\Mgrid} f(p_l -q)g(p_l).
\label{eq:smooth_func}
\end{align}

An e step slices $W_j$ from the 3D model $\Wgrid$ as follows:
\begin{align}
  \label{eq:estep}
  W_{ij} &= \sum_{l = 1}^{\Mgrid} f(p_l-R_jq_i) \Wgrid_{l}.
\end{align}

The C step inverses the interpolation of the e step by inserting the 2D slices 
back into the 3D grid,
\begin{align}
 \label{eq:cstep}
  \Wgrid_{l} &= \frac{\sum_{i = 1}^{\Mpix} \sum_{j = 1}^{\Mrot}
    f(p_l-R_jq_i)W_{ij}}{\sum_{i = 1}^{\Mpix} \sum_{j =
      1}^{\Mrot}  f(p_l-R_jq_i)}.
\end{align} 

After the C step in each iteration, the EMC algorithm checks the
following stopping criterion:
\begin{align}
\sum_l^{\Mgrid}|\Wgrid^{(n+1)}_{l=1} - \Wgrid^{(n)}_l | \le \epsilon,
\label{eq:EMC_exit}
\end{align}
where $\epsilon$ is a small positive number (we put $\epsilon = 0.001$
in practice in the experiments below).

We next explain the E and M steps in some detail. With
i.i.d.~diffraction patterns $K = (K_k)^{\Mdata}_{k=1}$, the
ML-estimator is given formally by
\begin{align} 
\hat{W} &= \arg\max_{W} \Mdata^{-1}\sum_{k = 1}^{\Mdata} \log
\Prob(K_{k}| W),
\label{eq:ML-estimator}
\end{align} 
 that is, estimated 2D slices are found by maximizing the likelihood of the  	
 diffraction patterns  for some probabilistic intensity model. 

Two factors make the optimization problem \eqref{eq:ML-estimator}
incomplete: i) the diffraction pattern $K_k$ cannot be directly
inserted back into a 3D volume due to the true rotation $R_k$ being
unknown; ii) the fluence $\phi_k$ of the $k$th diffraction pattern
$K_k$ is also unknown. To fix these two factors, we consider the
following ML-estimator instead,
\begin{align} 
\hat{W} &= \arg\max_{W} \Mdata^{-1}\sum_{k = 1}^{\Mdata}  \sum_{j = 1}^{\Mrot} \log
\Prob(K_{k}| W,R,\phi).
\label{eq:ML-estimator2}
\end{align} 

The original EMC algorithm \cite{EMC} assumed that the $i$th pixel of
the $k$th measured diffraction pattern $K_{ik}$ is Poissonian around
the unknown Fourier intensity $W_{ij}$,
\begin{align}
\Prob(K_{ik} = \kappa | W_{ij},R_j) = \prod_{j=1}^{\Mrot} \dfrac{(W_{ij})^\kappa e^{-W_{ij}}}{\kappa!}.
\end{align}
Some attempts \cite{EMC2, 3d_mimi} have been made to better take the
photon fluence into account by approximating the Poisson distribution
by a Gaussian distribution for high-intensity FXI diffraction
patterns, which unfortunately makes \eqref{eq:ML-estimator2}
nonlinear.

In this paper, instead of solving a scaled Poissonian probability
model directly, we propose a solution within the EMC framework using
ideas borrowed from non-negative matrix factorization (NNMF). More
precisely, we assume that the measured intensity of the $i$th pixel in
the $k$th diffraction pattern is Poissonian around the scaled unknown
Fourier intensity $\phi_{jk} W_{ij}$, i.e.~,
\begin{align}
\log \Prob(K_{ik}|W_{ij},R_j,\phi_{jk}) \propto K_{ik} \log W_{ij} + \log \phi_{jk} -\phi_{jk} W_{ij} := Q_{ijk}.
\end{align}

Summing over $i$, we obtain the joint log-likelihood function,
\begin{align}
Q_{jk} := \sum_{i=1}^{\Mpix} \left( K_{ik} \log W_{ij} + \log \phi_{jk} -\phi_{jk} W_{ij}\right).
\label{eq:ac_loglikelihood}
\end{align}

In the E step, we assume that the 2D slices $W$ and their fluences
$\phi$ are known, so that the rotational probability is explicitly
available by integrating the joint log-likelihood function
\eqref{eq:ac_loglikelihood} over the rotational space at the $(n+1)$th
iteration.
\begin{align}
  \nonumber
  P_{jk}^{n+1} =& P_{jk}^{n+1}(W^{n},\phi^{n})
  =: \Prob(R_j|K_k,\phi^{n}, W^{n}) \\
  =& \dfrac{w_j \exp (Q_{jk}(W^{n}))}
  {\sum_{j'=1}^{\Mrot} w_{j'} \exp (Q_{j'k}(W^{n}))}.
  \label{eq:E-step}
\end{align}

The M step freezes the rotational probability $P_{jk}$ at the
$(n+1)$th iteration, so that $\phi$ and $W$ may be obtained as
solutions to the following optimization problem:
\begin{align}
  \arg \max_{\phi,W} \sum_{ijk} \left(P_{jk} K_{ik} \log(\phi_{jk} W_{ij})
  - P_{jk} \phi_{jk} W_{ij} \right).
  \label{eq:Mlog}
\end{align}

We propose to solve for $\phi$ and $W$ jointly by directly translating
the optimization problem into an NNMF problem in the form of
minimizing the \emph{Klein divergence},
\begin{align}
  \nonumber
  \min_{\phi , W} D(P K || P \phi W) = & \min_{\phi, W} \sum_{ijk} \left( P_{jk} K_{ik} \log\frac{P_{jk} K_{ik}}{P_{jk} \phi_{jk} W_{ij}} - P_{jk} K_{ik} + P_{jk} \phi_{jk} W_{ij} \right)\\
  =& \min_{\phi , W} \left[ - \sum_{ijk} \left( P_{jk} K_{ik} \log(\phi_{jk} W_{ij}) - P_{jk} \phi_{jk} W_{ij} \right) + C \right],
  \label{eq:KL}
\end{align}
where $C$ is a constant.

The convergence of the NNMF algorithm is well-studied \cite{Lee2001,
  Zhi2011}, and the approach has been used successfully in
applications \cite{Lee1999, Dikmen2012,Yanez2014}. We minimize the
Klein divergence \eqref{eq:KL} via the multiplicative update rules
\eqref{Mstep:phi} and \eqref{Mstep:W}, which guarantees that
successive iterates of the Klein divergence is non-increasing.
\begin{align}
 \phi_{jk}^{(n+1)} &= 
  \dfrac{\sum_iK_{ik}}{\sum_i W_{ij}^{(n)}}   \dfrac{\sum_l \Wgrid_l^{(n-1)}}{  \sum_l \Wgrid_l^{(n)}},
  \label{Mstep:phi}
\end{align}
\begin{align}
  W^{(n+1)}_{ij} &= \frac{\sum_{k = 1}^{\Mdata}
    P_{jk}^{(n+1)}K_{ik}}
  {\sum_{k = 1}^{\Mdata}P_{jk}^{(n+1)}\phi_{jk}^{(n+1)}}, 
  \label{Mstep:W}
\end{align}
where $\sum_l \Wgrid_l^{(n-1)} / \sum_l \Wgrid_l^{(n)}$ is a normalization 
term.


\section{Uncertainty Analysis and Bootstrap Estimation}
\label{sec:boot}

In order to understand the overall uncertainty of the reconstruction
procedure in Fourier space, we investigate the successive steps of the
EMC algorithm. Armed with insights from this analysis, we suggest
practical bootstrap procedures to assess the limits of the
reconstruction resolution.

\subsection{Sources of uncertainty}
\label{subsec:sources}

To identify the sources of uncertainty, we work through the FXI
experiment setup and the EMC reconstruction procedure. On the one hand
the FXI experiment itself contributes several sources of errors: the
sample heterogeneity error due to inherent variations of biological
particles, the sample purity error due to there being a mixture of
different kinds of biological particles, and finally what may referred
to as an unexpected data error due to technical errors such as
detector malfunction, injector problems, and so on.

On the other hand, the EMC reconstruction procedure itself contributes
specific sources of errors or uncertainty: the smearing error $R_S$,
the rotational error $R_T$, the noise error $R_N$, and the fluence
error $R_F$. Currently, the errors related to FXI experimental
procedures are improving considerably \cite{detector,amoIn,injector},
and hence we only focus on the algorithmic errors and their
combinations. In summary,

\begin{description}
 \label{descp}
 \item [Smearing error $R_S$] This error is caused by a smearing
   effect in the compression step, and can in fact often be the
   dominating one. It can be reduced by using a finer model, or by
   higher-order interpolation methods.
   
 \item [Noise error $R_N$] This error is caused by noise in the
   diffraction patterns, and hence can be appreciated as a sampling
   error. It may also be caused by the data not filling the rotational
   space, i.e.~some voxels being empty or only having a small number
   of contributions due to very similar diffraction patterns being
   mapped to the same orientation, or simply because the number of
   diffraction patterns is too small.

 \item [Rotational error $R_T$] This error is due to the hidden data,
   i.e.~the unobserved particle rotation in the FXI experiments. $R_T$
   measures the error due to the rotational probability estimations in
   the E step \eqref{eq:E-step}.

 \item [The fluence error $R_{F}$] This error is due to the unobserved
   beam intensity at the object-beam interaction point in FXI
   experiments. The error is introduced when estimating the fluence in
   the M step \eqref{Mstep:phi}.
 \label{list:error}
\end{description}

Given these semantic definitions, we can now define these errors
mathematically, as well as discuss how to estimate them. We first
introduce two operators: $\oplus$ and $\circ$. The operator $\oplus$
is used when two or more errors are measured in the same estimation,
for example, $R_S \oplus R_N$ measures the effect of the smearing
error and the noise error at the same time. We also use the $\circ$
operator to connect each step of the EMC algorithm. For example, we
write the reconstruction $c\circ M(K^0,P^0,\phi^*) \circ E(K^0) \circ
e \circ \Wgrid$, when the EMC algorithm uses the noisy diffraction
patterns $K^0$, the 3D intensity $\Wgrid$, the correct fluence
$\phi^*$ and the estimated rotational probability $P^0$ in
computations.

In order to effectively speak of errors, we need to relate our results
against two reference 3D intensities: $\Wgrid^{*}$ and
$\Wgrid^{\perp}$. The reference $\Wgrid^{*}$ is the best possible EMC
reconstruction. In practice, it is obtained by inserting noiseless
diffraction patterns $K^*$ into their correct rotations, i.e.~applying
the compression step on the noiseless patterns given the correct
rotations, so $\Wgrid^* = c \circ K^*$. The reference $
\Wgrid^{\perp}$ is the 3D `truth' -- the 3D Fourier intensity without
any interpolation. $\Wgrid^{\perp}$ is used solely when the smearing
error $R_S$ is assessed.

Based on the set of noiseless diffraction patterns $K^*$, we define 3
additional sets of diffraction patterns: i) the nosiy diffraction
patterns $K^0 \sim \po(K^*)$, where $\po(K^*)$ represents Poisson
random variables with rate parameters (means) $K^*$; ii) the patterns
with randomly varying fluence $K^{f*} = \phi K^*$, with $\phi =
(\phi_k)_{k=1}^{\Mdata}$; iii) the corresponding noisy patterns
$K^{f0}\sim \po(K^{f*})$. The true rotational probability and fluence
are $P^*$ and $\phi^*$ respectively, while the estimated ones are
denoted by $P^0$ and $\phi^0$. Table~ \ref{tab:notations} summarizes
these notations.
\begin{table}[!htb]
\begin{tabular}{l l}\hline
$\Wgrid^{\perp}$ 	& The 3D `truth' \\
$\Wgrid^*$ 			& The best possible EMC reconstruction \\
$K^*$				& Noiseless diffraction patterns \\
$K^{0}$				& Noisy diffraction patterns\\
$K^{f*}$				& Diffraction patterns taking fluence into account\\
$K^{f0}$				& Corresponding noisy diffraction patterns \\ \hline
\end{tabular}
\caption{Notation for assessing algorithmic errors.}
\label{tab:notations}
\end{table}

We may now directly measure the algorithmic errors by comparing the
reference 3D intensities with the reconstructed intensities. These
measured errors are 3D maps, which can be projected to univariate
error measures using the error metrics discussed in \S
\ref{subsec:error_metric}. Table \ref{tab:error_cal} lists the
constructive definition of the algorithmic errors.

\begin{table}[!htb]
\centering
\begin{tabular}{l l r }\hline
Name & Error(s) & Definition  \\ \hline
Smearing & $R_S$ & $ \Wgrid^* - R\Wgrid^{\perp}$\\
Noise & $R_N$ & $ c \circ K^0 - R\Wgrid^*$  \\	
Rotational & $R_T$ & $c \circ M(K^*,P^0,\phi^*) \circ E(K^0) \circ e 
    			\circ \Wgrid - R\Wgrid^*$ \\ 	
 & $R_N \oplus R_T$ & $ c \circ M(K^{0},P^0,\phi^*) \circ E(K^0) \circ e \circ \Wgrid - R\Wgrid^*$ \\		
Fluence & $R_F$  & $ c\circ M(K^{f*},P^*,\phi^0) \circ E(K^{f0}) \circ e 
			\circ \Wgrid  - R\Wgrid^*$\\	 		
 & 	$R_F \oplus R_N$ & $ c \circ M(K^{f0},P^*,\phi^0) \circ E(K^{f0}) \circ e \circ \Wgrid - R\Wgrid^*$ \\			
  & 	$R_F \oplus R_T$ & $ c \circ M(K^{f*},P^0,\phi^0) \circ E(K^{f0}) \circ e \circ \Wgrid - R\Wgrid^*$ \\			
  & 	$R_F \oplus R_N \oplus R_T$ & $ c \circ M(K^{f0},P^0,\phi^0) \circ E(K^{f0}) \circ e \circ \Wgrid - R\Wgrid^*$ \\									 
\hline			  			
\end{tabular}
\caption{A constructive definition of algorithmic errors and their
  combinations. To subtract an estimate from a reference map, a
  rotation $R$ which takes them into the same frame of reference must
  always be performed. Note that, in order to measure the smearing
  error $R_S$ in combination with others, we simply use
  $\Wgrid^{\perp}$ instead of $\Wgrid^*$.}
\label{tab:error_cal}
\end{table}
\subsection{Bootstrap estimators}
\label{subsec:boot_theory}

The algorithmic errors as defined previously can be measured only when
a reference 3D intensity is known. For other situations, we now
develop practical bootstrapping procedures. Bootstrapping
\cite{bootstrapIntro,bootstrapJack} is a general computational
methodology that relies on random resampling of collected data. It is
used to estimate stability properties of an estimator, e.g.~its
variance and standard derivation. For the EMC algorithm, we introduce
two bootstrap schemes -- one `standard' approach based on common
practice and one approach specially designed for the EM framework.

\subsubsection{Standard bootstrap method}
\label{subsubsec:stdboot}

The standard bootstrap relies on resampling input diffraction patterns and 
reconstructing them using the EMC procedure. The workflow of the standard 
bootstrap method is illustrated in Figure~\ref{fig:boot_scheme} (left).

Let the diffraction patterns $K = (K_k)_{k=1}^{\Mdata}$ be the whole
bootstrap universe. The bootstrap replacement method generates $B$
bootstrap samples $ (S_r)_{r=1}^B$, and each sample $S_r$ contains
$\Mdata$ frames that are randomly chosen from $K$ with replacement. In
other words, every sample only contains a certain part of $K$,
including duplicate frames. The EMC algorithm then reconstructs each
sample yielding $(\Wgrid_r)_{r=1}^B$. The EMC algorithm is also used
to reconstruct the whole bootstrap universe $K$ yielding $\Wgrid_a $.


Once all reconstructions are obtained, the bootstrap mean is given by 
\begin{align}
  \Wgrid_M &\equiv \dfrac{1}{B}\sum_{r=1}^{B} R_{r}\Wgrid_r,
    \label{eq:stdboot_mean}
\end{align}
where $R_r$ is the rotation required to align $\Wgrid_r$ to
$\Wgrid_a$.  Consequently, $\Wgrid_M$ is also aligned to
$\Wgrid_a$. In practice, we determine the rotation $R_r$ by solving an
optimization problem, see~\eqref{align_norm}.

With \eqref{eq:stdboot_mean} defined, the bootstrap estimate of the
variance is defined as follows:
\begin{align}
  \mathbb{V} &= 
  \dfrac{1}{B-1}\sum_{r=1}^{B}( R_{r}\Wgrid_r-\Wgrid_M )^{2}.
  \label{eq:stdboot_variance}
\end{align}
The standard error of the mean is proportional to the square root of the 
variance,
\begin{align}
   R_{std} &\propto
  \sqrt {\dfrac{ \mathbb{V} }{B}}. 
    \label{eq:stdboot_std}
\end{align}

Since each bootstrap sample only sees a portion of the bootstrap universe, it 
may be biased. We estimate this bias by
\begin{align}
R_{bias} &= \Wgrid_M- \Wgrid_{a}.
    \label{eq:stdboot_bias}
\end{align}

Since the EMC algorithm uses the same grid for reconstructing all bootstrap 
samples, these reconstructions all have the same level of smearing error. This 
means that none of the bootstrap estimates we have introduced can reliably 
estimate the smearing error. Instead, we estimate it separately as follows:
\begin{align}
\hat R_{S} &= c \circ e \circ \Wgrid_M - \Wgrid_M,
\label{eq:stdboot_rs}
\end{align}
that is, we expand $\Wgrid_M$ into $\Mdata$ slices and then compress them back 
into the 3D volume. 
 
An estimator of the total reconstruction uncertainty $R_{total}$ can now be 
formed by adding the standard error, the estimated bias, and the estimated 
smearing error together,
\begin{align}
    R_{total}^{2} 
    &=  \beta^2 R_{std}^{2} + R_{bias}^{2} + \hat R_{S}^2,
  \label{eq:boot_val}
\end{align}
where $\beta$ is the constant for the proportionality in 
\eqref{eq:stdboot_std}. In practice, we take $\beta = 2$ or $3$.

The standard bootstrap procedure  for the EMC algorithm is summarized in 
Algorithm~\ref{alg:stardboot}. 

\begin{algorithm}[H]
  \begin{algorithmic}
    \STATE
    \STATE \textbf{Input:} \textnormal{Initial guess of the 3D
      intensity $\Wgrid^{(0)}$  and the bootstrap universe of diffraction patterns $K$.}
    \STATE \textbf{Output:} \textnormal{ Bootstrap mean $\Wgrid_M$ together with an estimated uncertainty $R_{total}$.}
  \end{algorithmic}
  \begin{algorithmic}[1]  
  	\STATE Run the EMC algorithm on the bootstrap universe $K$, yielding $\Wgrid_a$.
    \STATE Generate bootstrap samples $(S_r)_{r=1}^B$ by resampling  with replacement in the bootstrap universe $K$.
    \FOR{ $r = 1,\ldots,B$}
	\STATE Run the EMC algorithm on the bootstrap sample $S_r$  until \eqref{eq:EMC_exit} is satisfied, yielding $\Wgrid_r$.
    \ENDFOR
    \STATE Compute the standard error $R_{std}$  and the bootstrap sample bias $R_{bias}$ via \eqref{eq:stdboot_std} and  \eqref{eq:stdboot_bias} respectively.
    \STATE Calculate the estimated smearing error $\hat R_{S}$ via \eqref{eq:stdboot_rs}.
    \STATE Estimate the total reconstruction uncertainty $R_{total}$ by \eqref{eq:boot_val}.
  \end{algorithmic}
  \caption{The standard bootstrap method for the EMC algorithm.}
  \label{alg:stardboot}
\end{algorithm}

\subsubsection{The EM algorithm with bootstrapping (EMB)}
\label{subsubsec:EMB}

\begin{figure}[!htb]
  \subfloat{\includegraphics[width=.5\textwidth]{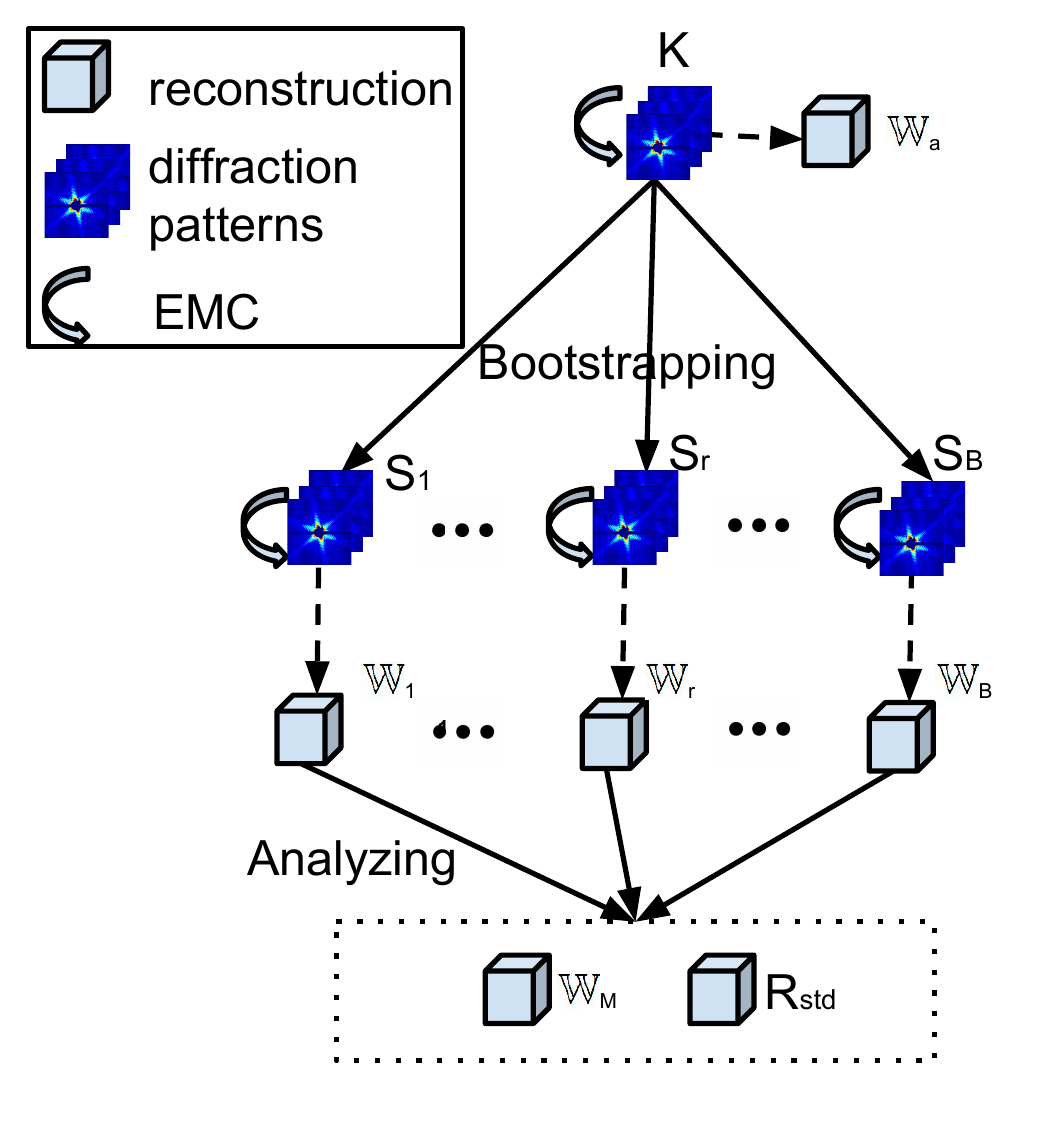}}
  \subfloat{\includegraphics[width=.5\textwidth]{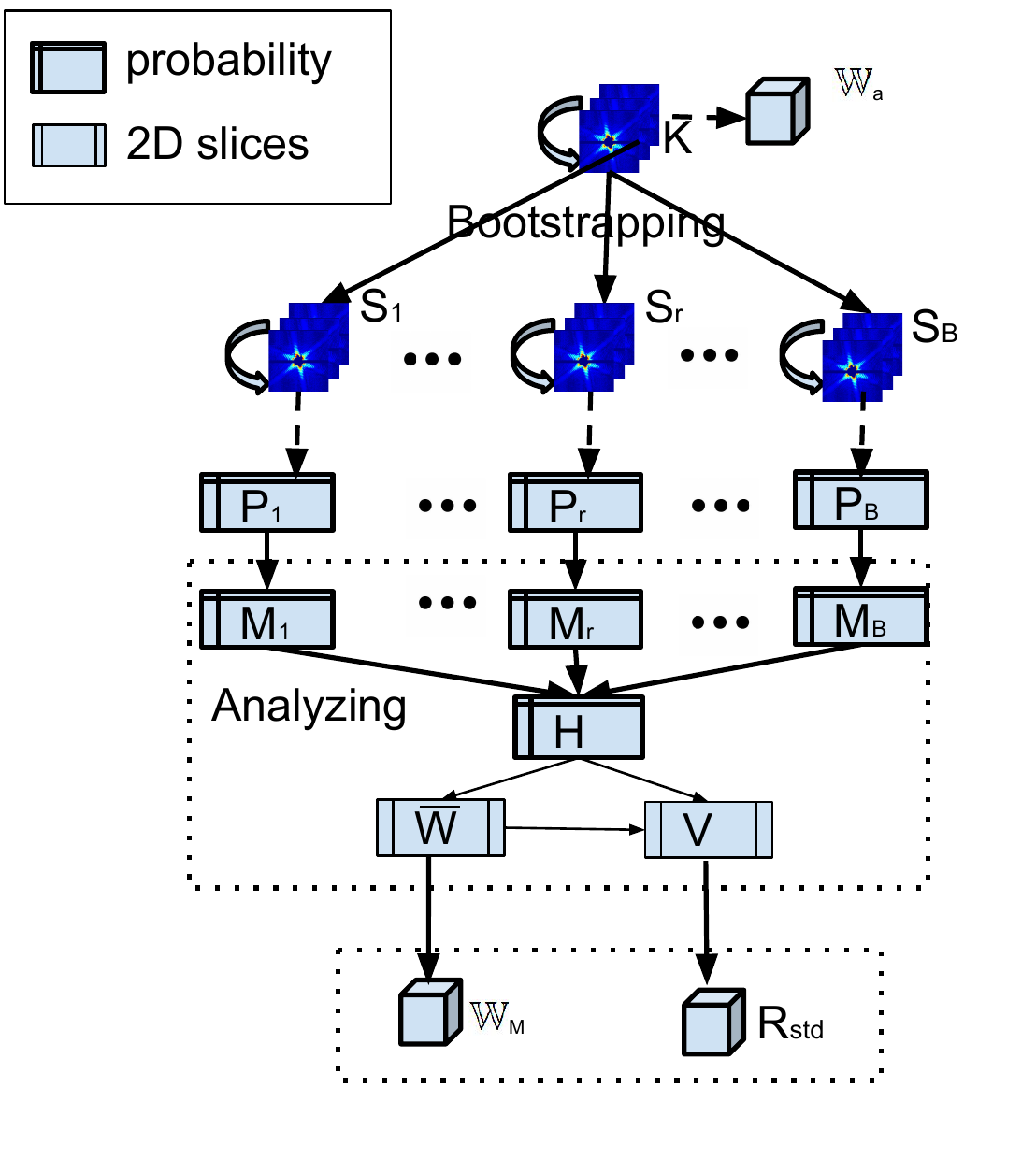}}\\
  \caption{Bootstrap schemes for EMC: the standard bootstrap
    (\emph{left}) and the EMB (\emph{right}).}
  \label{fig:boot_scheme}
\end{figure}

The EMB is a general method that applies bootstrapping under the EM
framework \citep{emb1,emb2}. Similar to the standard bootstrap method,
the EMB method also relies on random resampling. However, instead of
analyzing the final 3D model, the EMB method calculates a bootstrap
mean of probabilities.  This calculation can be done at every
iteration \cite{emb2} or after all reconstructions have finished
\cite{emb1}. Here we use the later method, since it can work together
with the standard bootstrap method by only adding a small amount of
computations. Figure \ref{fig:boot_scheme} (right) illustrates the
workflow of the EMB method.

We now explain the EMB procedure in some detail. The EMC algorithm
first runs on the whole bootstrap universe $K$, yielding $\Wgrid_a$,
and saves the estimated fluence $\bar \phi$ at the final
iteration. The EMB method then generates the bootstrap sample
$(S_r)_{r=1}^B$ using the same resampling method as in the standard
bootstrap method described in \S \ref{subsubsec:stdboot}. For all
bootstrap samples, the EMC algorithm executes until it meets the
stopping criterion \eqref{eq:EMC_exit}, and saves the estimated
rotational probabilities $(P_{jkr})_{r=1}^{B}$ for each bootstrap
sample $ (S_r)_{r=1}^B$. Then the EMB method picks out the \emph{mode}
(i.e.~the most probable rotation) $(M_{jkr})_{r=1}^B$ for each frame
in the bootstrap sample,
\begin{align}
   M_{jkr}=
   \begin{cases}
   1 &\mbox{if}\; P_{jk'r} = \max_j P_{jk'r}\\
   0 &\mbox{otherwise}
   \end{cases}.
   \label{eq:EMB_M}
\end{align}
where the $k'$th frame of the $r$th bootstrap sample is the $k$th frame of the 
bootstrap universe $K$.

Following this step, EMB combines all the modes, so that each frame $K_k$ now 
comes equipped with an empirical distribution over the rotational space. 
The bootstrap mean of those modes is 
\begin{align}
H_{jk} = \dfrac{1}{B}\sum_{r=1}^{B}  M_{jkr}.
   \label{eq:EMB_H}
\end{align}

Using this empirical distribution, the EMB method then computes the 2D 
bootstrap mean as follows:
\begin{align}
\bar W_{ij} = \dfrac{\sum_k H_{jk} K_{ik}}{ \sum_k H_{jk}  \bar \phi_{jk}},
   \label{eq:EMB_W}
\end{align}
where $\bar \phi$ is the estimated fluence at the final iteration when 
reconstructing the bootstrap universe $K$.

The 2D bootstrap variance is also defined
\begin{align}
\bar V_{ij}= \dfrac{\sum_k H_{jk} (K_{ik}-\bar \phi_{jk} \bar W_{ij})^2}{ \sum_k H_{jk}}.
   \label{eq:EMB_V}
\end{align}

To generate a comparable result to the standard bootstrap method, the EMB next 
compresses the bootstrap mean $\bar W$ and variance $\bar V$ by 
\eqref{eq:cstep}, yielding a 3D mean $\Wgrid_M$ and a 3D variance $\mathbb{V}$, 
respectively. The 3D standard error $R_{std}$ is again proportional to the 
square root of the variance \eqref{eq:stdboot_std}.

Once all reconstructions have been obtained, the EMB method calculates the 
bootstrap sample bias via 
\begin{align}
R_{bias} &= \Wgrid_M- R\Wgrid_{a},
    \label{eq:EMB_bias}
\end{align}
where $R$ is again the required rotation to align $\Wgrid_a$ to $\Wgrid_{M}$.

After using \eqref{eq:stdboot_rs} to estimate the smearing error $\hat R_S$, 
the EMB method estimates the total reconstruction error $R_{total}$ again via 
\eqref{eq:boot_val}.

The EMB method is summarized in Algorithm~\ref{alg:EMB}.

\begin{algorithm}[H]
  \begin{algorithmic}
	\STATE
    \STATE \textbf{Input:} \textnormal{Initial guess of the 3D
      intensity, $\Wgrid^{(0)}$, and the bootstrap universe of diffraction patterns $K$.}
    \STATE \textbf{Output:} \textnormal{ Bootstrap mean $\Wgrid_M$ together with an estimated uncertainty $R_{total}$.}
  \end{algorithmic}
  \begin{algorithmic}[1]  
  	\STATE Run the EMC algorithm on the bootstrap universe $K$, yielding $\Wgrid_a$ and the estimated fluence at the final iteration $\bar \phi$. 
    \STATE Generate bootstrap samples $(S_r)_{r=1}^B$ by resampling  with replacement in the bootstrap universe $K$.
    \FOR{ $r = 1,\ldots,B$}
	\STATE Run the EMC algorithm on the bootstrap sample $S_r$  until \eqref{eq:EMC_exit} is satisfied, and save the probability $P_{jkr}$ at the final iteration.
    \ENDFOR  
    \STATE Compute the modes  and the empirical distribution via \eqref{eq:EMB_M} and \eqref{eq:EMB_H} respectively.
    \STATE Calculate the 2D  mean and the 2D  variance by \eqref{eq:EMB_W} and \eqref{eq:EMB_V} respectively.
    \STATE Assemble the 2D mean and the 2D variance back into 3D volumes via \eqref{eq:cstep}, yielding the 3D mean $\Wgrid_M$ and the 3D variance $\mathbb{V}$.
    \STATE Compute the bootstrap sample bias $R_{bias}$ by \eqref{eq:EMB_bias} and the standard error $R_{std}$ via \eqref{eq:stdboot_std}.
    \STATE Calculate the estimated smearing error $\hat R_{S}$ via \eqref{eq:stdboot_rs}.
    \STATE Estimate the total reconstruction uncertainty $R_{total}$ by \eqref{eq:boot_val}.
  \end{algorithmic}
  \caption{The EMB method.}
  \label{alg:EMB}
\end{algorithm}

With these two bootstrap schemes, we hope to accurately estimate the 
algorithmic errors by reasoning essentially as in 
\begin{align}
\label{eq:trangle_in}
\|R_{total}\|  &\approx \|R_S\| + \|R_T\| + \|R_N\| + \|R_F\| \\
&\gtrsim 
\nonumber
\|R_S \oplus R_T \oplus R_N \oplus R_F \|.
\end{align}
However, the nonlinear interaction between the various sources of
uncertainty may in fact imply that $\|R_S\| + \|R_T\| + \|R_N\| +
\|R_F\| < \|R_S \oplus R_T \oplus R_N \oplus R_F \|$. We usually
expect that \eqref{eq:trangle_in} is a robust estimate of the overall
reconstruction uncertainty, at least when reconstructing a
sufficiently large set of diffraction patterns.


\section{Experiments}
\label{sec:exp}

We now proceed to measure some actual algorithmic errors and assess
the sharpness of our bootstrap methodology when confronted with
synthetic data. In \S\ref{subsec:setup} we detail our experimental
setup and in \S\ref{subsec:error_metric} we discuss the process of
estimating the errors defined in
\S\ref{sec:boot}. \S\ref{subsec:exp_errors} is devoted to an
investigation of the algorithmic errors and their
combinations. Finally, the sharpness and robustness of the
bootstrapping procedures are investigated in
\S\ref{subsec:sharpness}--\ref{subsec:robustness}.

To reduce the computing time, we used our data distribution scheme
described in \cite{hpcEMC} for parallelization. All implementations
were compiled with GCC 4.4.7, CUDA 7.5, and Open MPI 1.8.1. With
respect to the hardware, we used a cluster with 4 Nvidia Kepler GPUs
in each node, interconnected via an InfiniBand 32Gbit/s fabric.

\subsection{Setup and synthetic data}
\label{subsec:setup}

As summarized in \S\ref{subsec:scattering}, we know that a diffraction pattern 
is a central symmetric image containing interference of waves. The 
interference pattern is dependent on the rotation and shape of the target 
particle. To be able to discuss reproducible reconstructions, we propose the 
following 3D synthetic model of a 3D diffraction pattern,
 \begin{align}
 \centering
 M(\alpha,\beta,k) &= C  \sin^2( R(\alpha)/2)R(\alpha)^{k} + C \sin^2( R(\beta)/2)R(\beta)^{k},
 \label{eq:M2}
 \end{align}
\begin{align}
 \centering
 R(\alpha) = \sqrt{\alpha_0  X^2+ \alpha_1 Y^2+ \alpha_2 Z^2},
 \label{eq:Rdist}
\end{align}
where $(X,Y,Z)$ are 3D meshgrid coordinates whose origin is the center
of a $64^3$ cube, where 3 different grids were used in our
experiments, $\Mgrid = [64^3, 128^3, 256^3]$, dividing the coordinates
$(X,Y,Z)$ with $[1,2,4]$, respectively. Further, $k$ is the intensity
drop exponent, $C$ the intensity constant, and $\alpha =
(\alpha_0,\alpha_1,\alpha_2)$ and $\beta = (\beta_0,\beta_1,\beta_2)$
are shape vectors.
 
For the numerical experiments in this paper, our 3D `truth' is $\Wgrid^{\perp} 
:= M(\alpha = [1.5,0.3,0.5], \beta=[0.2,0.9,1],k=-4)$. We also randomly and 
uniformly picked up $\Mdata = 1000$ or $5000$ rotations from 400,200 rotations 
sampled from the 600 Cell \cite[Appendix C]{EMC}. With the selected rotations, 
we generated $\Mdata$ noiseless diffraction patterns $K^*$ from $\Wgrid^{\perp}
$ via the expansion step \eqref{eq:estep}. Using $K^*$, we also generated 
patterns sampled as a Poissonian signal $K^0 \sim \po(K^*)$, patterns with 
randomly varying fluence $K^{f*} = \phi K^*$, and the corresponding Poissonian 
patterns $K^{f0} \sim \po(\phi K^*)$. Here, the fluence $\phi$ was uniformly 
and randomly chosen in $(0.9,1.2)$. All these parameters were chosen to 
reasonably mimic realistic conditions \cite{Bozek2009,EMC}. 
  
\begin{figure}[!htb]
    \includegraphics[width=0.2\textwidth]{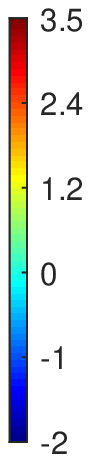}\hspace{0em}
   \includegraphics[width=0.78\textwidth]{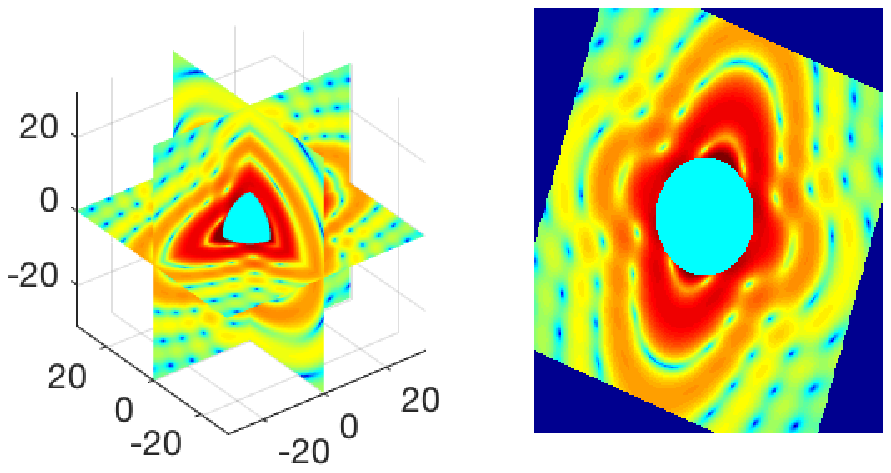}\\ 
   \caption{\emph{Left:} Slices view of $\Wgrid^{\perp} := M(\alpha =
     [1.5,0.3,0.5], \beta=[0.2,0.9,1],k=-4)$, as defined in
     \eqref{eq:M2}. The diameter of the missing data region are
     $[16,32,64]$ voxels at $\Mgrid= [64^3,128^3,256^3]$.
     \emph{Right:} A noiseless synthetic diffraction pattern generated
     from $\Wgrid^{\perp}$. Both figures are drawn in logarithmic
     scale.}
   \label{fig:syn_model}
\end{figure}

In FXI experiments,  a hole is normally located in the middle of the detector 
to let the unscattered X-ray photons pass, and consequentially a missing data 
area exists in the middle of all diffraction patterns. To make our synthetic 
diffraction patterns realistic, we also mask out a circular zero region, with 
radii $[8,16,32]$ pixels at the respective diffraction pattern sizes $
\Mpix=[64^2,128^2,256^2]$. Figure~\ref{fig:syn_model} shows the 3D `truth' $
\Wgrid^{\perp}$ with a central missing data region and a noiseless diffraction 
pattern.
 
\subsection{Error metrics}
\label{subsec:error_metric}
Since it is reasonable to compare Fourier intensities about the same frequency, 
we propose a simple method to compare two 3D intensities in radial shells as 
follows.

Let $S = (S_u)_{u=1}^{U}$ be the selected radial shells of a 3D intensity. The 
$u$th shell is given by $S_{u} =\{s = (x,y,z); s_{u} \le \|s\| < s_{u+1}\}$, 
where $s$ is a point (voxel) at position $(x,y,z)$, and $\|s\|$ is the
Euclidean norm.

We then define the (strong) error of the $u$th shell as follows:
\begin{align}
  \hat{e}_{u} (\Wgrid_1,\Wgrid_2) &= \frac{1}{|S_{u}|} \sum_{s \in S_{u}} \frac{
    |(\Wgrid_1)_s - (R\Wgrid_2)_s|}{\max(\rho,(|\Wgrid_1|_s + |R\Wgrid_2|_s)/2)},
\label{eq:strong_error}
\end{align}
 where $\rho$ is a small cutoff number to prevent dividing by zero, and where
 $R$ is the rotation required to align $\Wgrid_2$ to $\Wgrid_1$. This error  
 metric is a strong and very revealing measure, since it effectively compares
 relative errors in every point of $\Wgrid_1$ and $\Wgrid_2$. Alternatively, we 
 consider a weaker version which rather compares each shell in an average sense 
 only,
\begin{align}
  e_u(\Wgrid_1,\Wgrid_2) &= \frac{\sum_{s \in S_{u}}
    |(\Wgrid_1)_s - (R\Wgrid_2)_s|}{\sum_{s \in S_{u}} |(\Wgrid_1 + R\Wgrid_2)_s|/2}.
      \label{eq:weak_error}
\end{align}

In turn, we align $\Wgrid_1$ to $\Wgrid_2$ by solving the following
optimization problem.
\begin{align}
  \arg \min_{R} \quad U^{-1}  \sum_{u=1}^{U}  {e}_u (\Wgrid_1 ,R\Wgrid_2),
    \label{align_norm}
\end{align}
that is, we find a proper alignment by minimizing the total weak error
using a global optimization algorithm \cite{optGlobal}. To be more
robust, it is sometimes useful to align $\Wgrid_1 $ and $\Wgrid_2$
several times from different start rotations, and pick up the mode of
this sample. In practise, this minimization problem was never a major
obstacle in our experiments.

In order to get a baseline for these error metrics, we now explore two
basic error measurements: the 100\% and the 50\% hidden-data errors.
Let $\Wgrid^{\times}$ be a reconstruction produced by inserting
$\Mdata = 1000$ \emph{noiseless} diffraction patterns \emph{randomly}
into a 3D volume.  Similarly, the reconstruction
$\Wgrid^{\frac{1}{2}\times}$ is obtained by inserting the first half
of those noiseless patterns randomly into a 3D volume, and the rest
into the \emph{correct} rotations. Comparing $ \Wgrid^{\times}$ and
$\Wgrid^{\frac{1}{2}\times}$ with the 3D `truth' $ \Wgrid^{\perp}$
defines 4 errors: the strong and the weak 100\% errors $\hat{R} _{100}
= \hat{e}(\Wgrid^{\times},\Wgrid^{\perp})$, $R_{100} =
e(\Wgrid^{\times},\Wgrid^{\perp})$, and the strong and the weak 50\%
errors $ \hat{R}_{50}=\hat{e}
(\Wgrid^{\frac{1}{2}\times},\Wgrid^{\perp})$ and $R_{50} =
e(\Wgrid^{\frac{1}{2}\times},\Wgrid^{\perp})$.

\begin{figure}[!htb]
  \subfloat{\includegraphics[width=.5\textwidth]{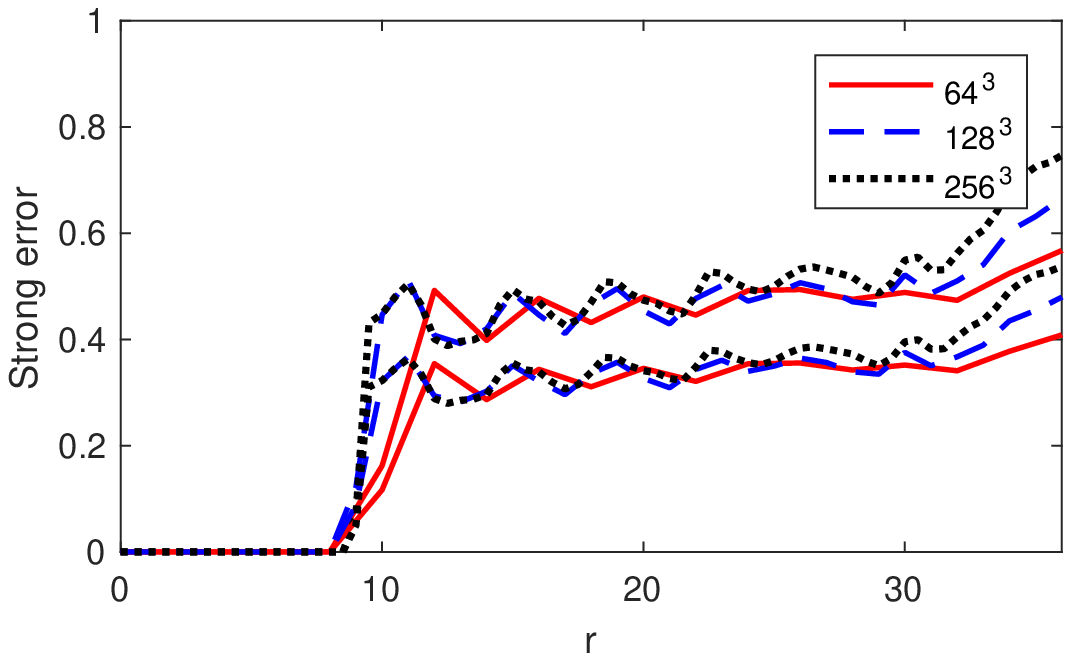}}
  \subfloat{\includegraphics[width=.5\textwidth]{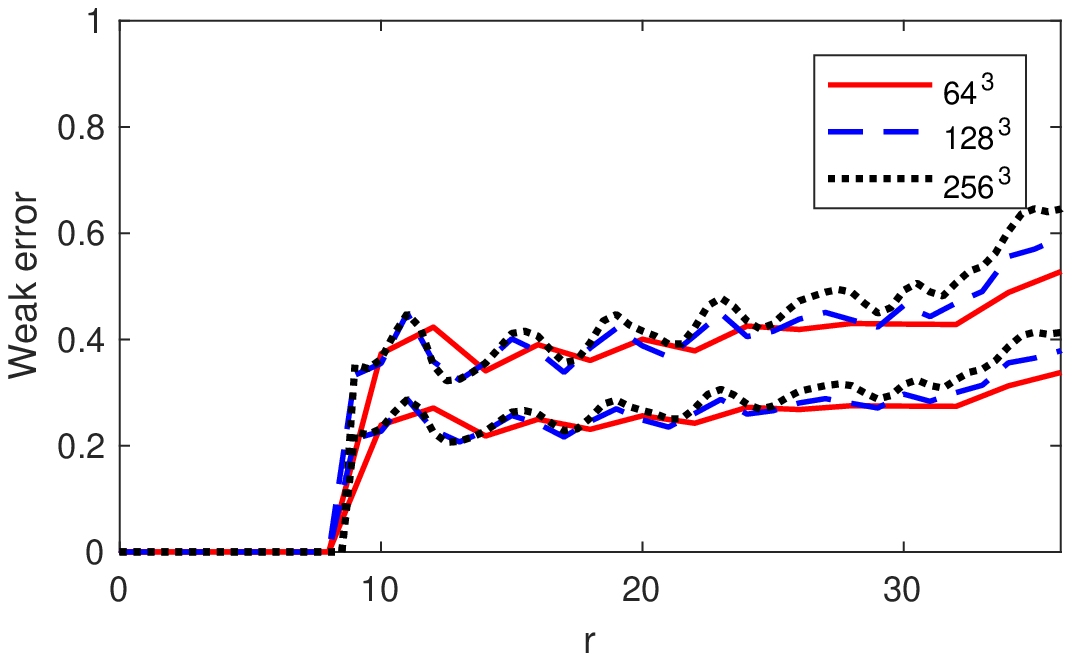}}\\
  \caption{ The hidden data errors at different model sizes.  
  \emph{Left}: the top three line are $\hat{R}_{100}$, and the bottom three
    are  $\hat{R}_{50}$. \emph{Right}: corresponding plots for the weak errors
    $R_{100}$ and $R_{50}$.}
\label{fig:R_rand}
\end{figure}

Figure~\ref{fig:R_rand} shows the strong and weak 100\% and 50\%
hidden-data errors for the synthetic data model $\Wgrid^{\perp} :=
M(\alpha = [1.5,0.3,0.5], \beta=[0.2,0.9,1],k=-4)$, from
Figure~\ref{fig:syn_model}. As can be seen, the strong errors
($\hat{R}_{100}, \hat{R} _{50}$) are larger than their weak
counterparts ($R_{100}, R_{50}$), but all show the same trend. The
errors rise faster at the shell distance $r :=||s|| \in (30,32)$, due
to the truncated domain - the regions filled with zeros on the corners
of the diffraction patterns.

Based on the above, we may state that a reconstruction procedure
\emph{fails} if the reconstruction uncertainty is larger than
$R_{100}$, or approximately when the means of the strong or,
respectively, the weak errors for $r \in (8,30)$ are larger than 0.50
and 0.37. We may also claim that a \emph{proper} reconstruction should
generally have uncertainties less than $R_{50}$, or that the means of
the strong and the weak error for $r \in (8,30)$ should be less than
0.32 and 0.23, respectively.

\subsection{Influences of errors}
\label{subsec:exp_errors}

In this section we investigate the algorithmic errors as defined in \S
\ref{subsec:sources}. Recall that the 3D `truth' $\Wgrid^{\perp}$ and the EMC 
best reconstruction $\Wgrid^*$ are both used when measuring the algorithmic 
errors. 

\subsubsection{The smearing error $R_S$}

We first measured the error that is induced by the compression step,
i.e. the smearing error $R_S$. As defined in \ref{tab:error_cal},
$R_S$ compares the EMC best reconstruction $\Wgrid^*$ with the 3D
`truth' $\Wgrid^{\perp}$. We measured this error in both the strong
sense $\hat{e}(\Wgrid^*,R\Wgrid^{\perp})$ and the weak sense
$e(\Wgrid^*,R\Wgrid^{\perp})$, where $\hat{e}_k$ and $e_k$ are defined
in \eqref{eq:strong_error} and \eqref{eq:weak_error}, respectively.

\begin{figure}[!htb]
  \subfloat{\includegraphics[width=.5\textwidth]{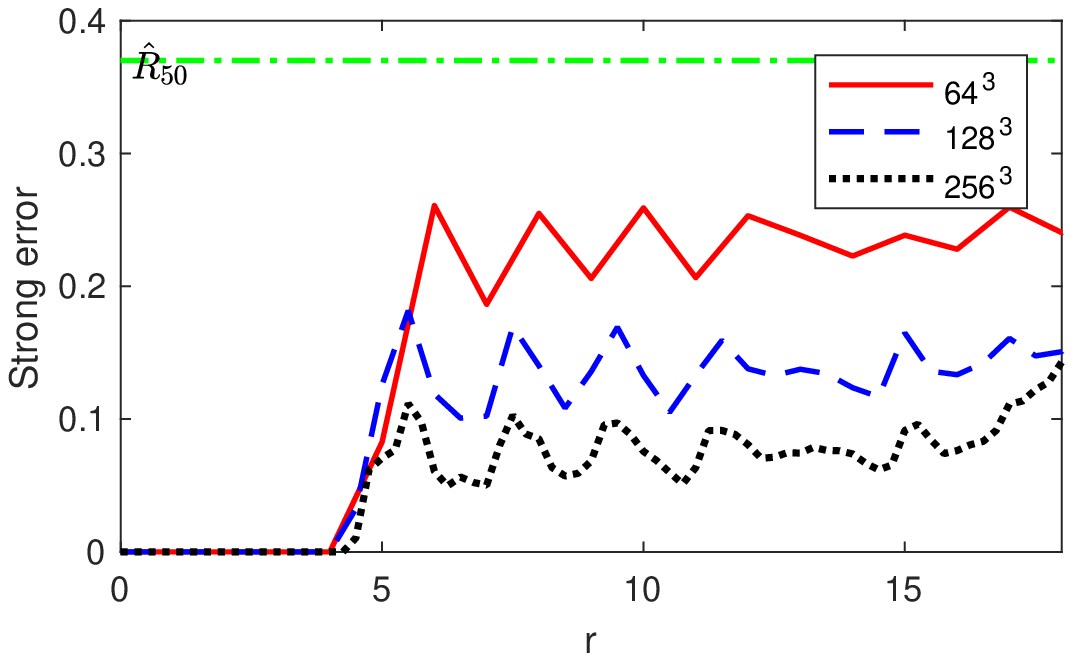}}
  \subfloat{\includegraphics[width=.5\textwidth]{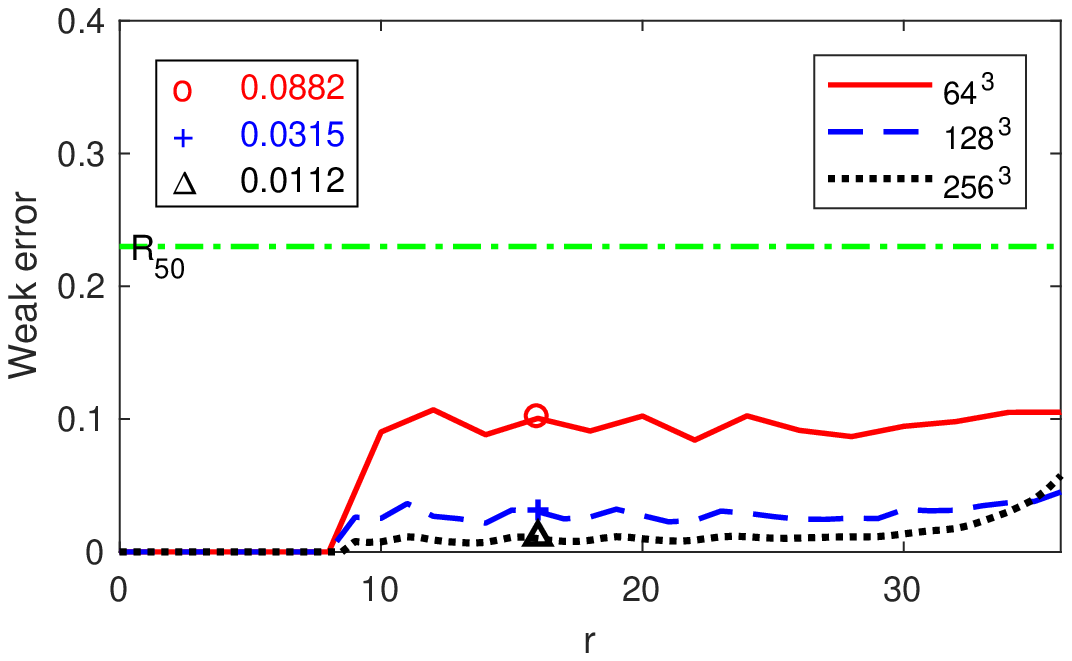}}\\ 
  \caption{ The strong (\emph{left}) and the weak (\emph{right})  smearing  
  errors. The dash-dot lines are the average $R_{50}$ and $\hat{R}_{50}$, 
  respectively.}
\label{fig:RS}
\end{figure}

Figure~\ref{fig:RS} shows these errors at the grid sizes $\Mgrid =
[64^3,128^3,256^3]$. As expected, the strong error is larger and more
sensitive than the weak error, but both error definitions follow a
similar trend. Since linear interpolation is used for implementing
both the expansion \eqref{eq:estep} and the compression
\eqref{eq:cstep} step, we expect and observe an overall typical
$O(h^2)$ smearing error, hence $R_S$ can be reduced by a factor of
four by doubling the side length of the grid. Further, the $64^3$
resolution performs bad -- the weak $R_S$ error is around $\frac{1}{2}
R_{50}$, due to strong aliasing artifacts in the diffraction patterns.

Since the strong and the weak errors performed similarly, from now on
we only present the weak error $e_k$ together with the average
${R}_{50}$ as a reference.

\subsubsection{The hidden data and the noise error ($R_T$ and  $R_N$).}

We also studied the error that is induced by estimating the unobserved
particles rotations - the rotational error $R_T$, and the error caused
by noise in data - the noise error $R_N$.
  
\begin{figure}[!htb]
  \subfloat{\includegraphics[width=.5\textwidth]{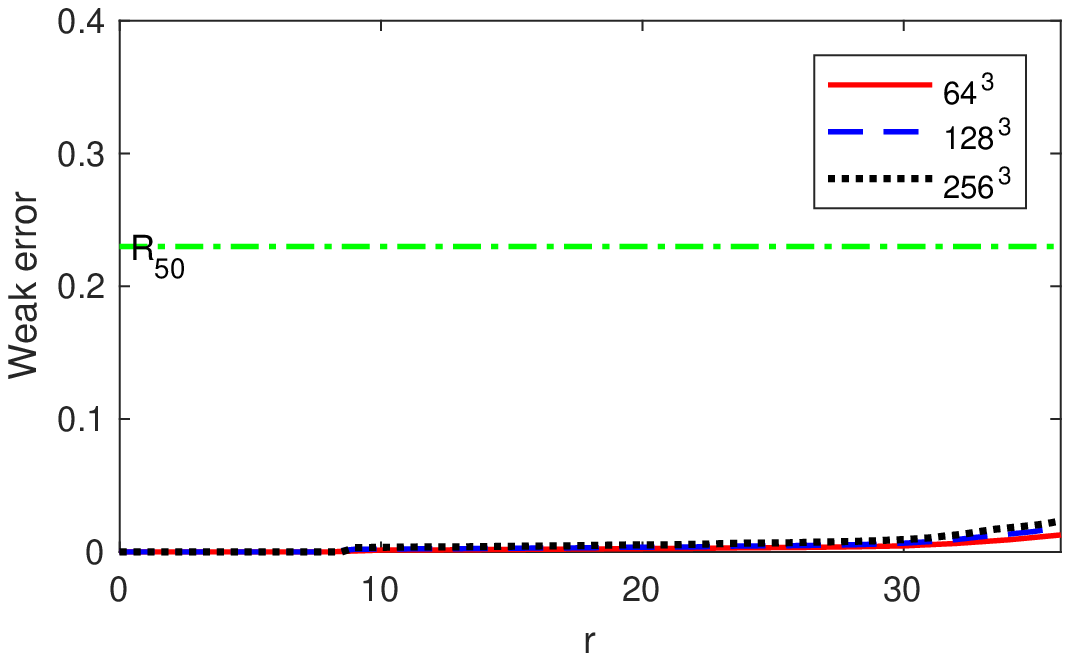}} 
  \subfloat{\includegraphics[width=.5\textwidth]{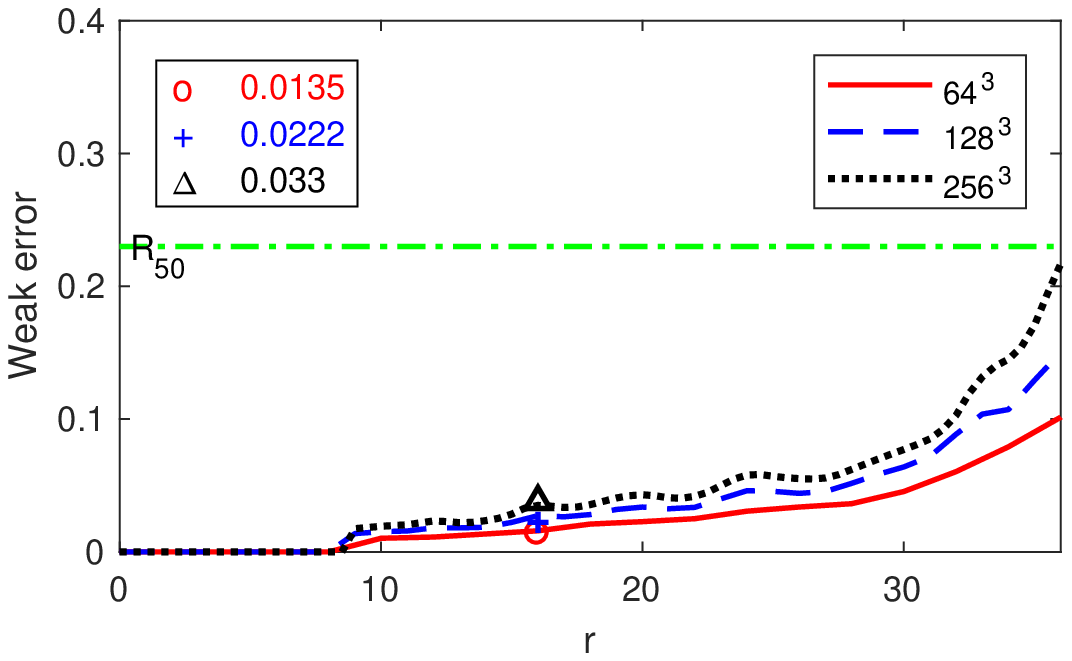}}\\ 
  \subfloat{\includegraphics[width=.5\textwidth]{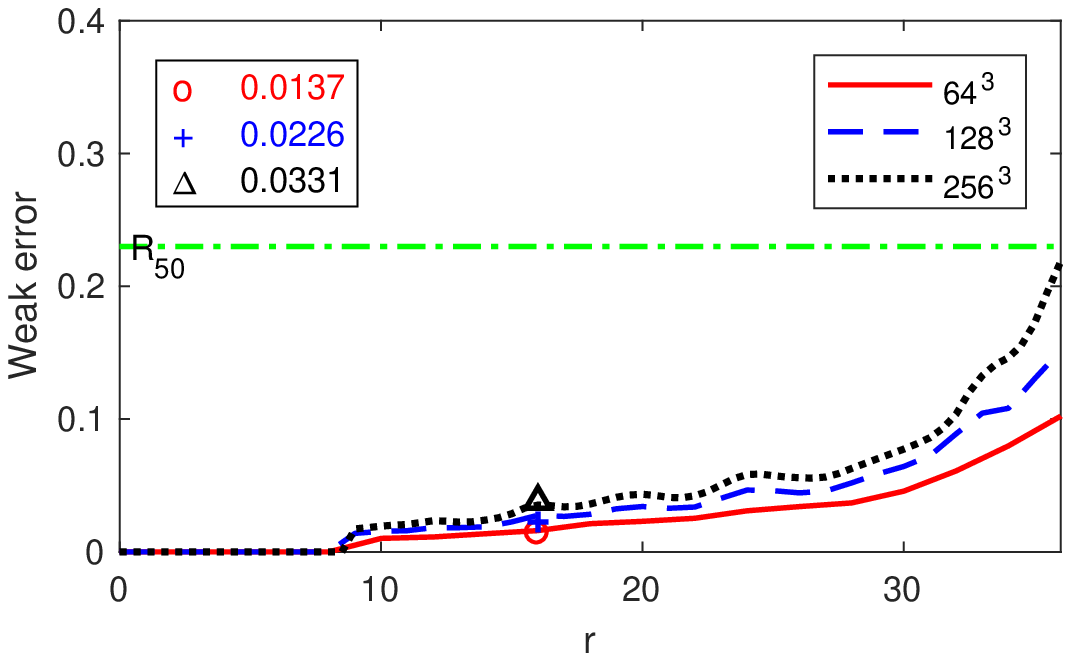}}
  \subfloat{\includegraphics[width=.5\textwidth]{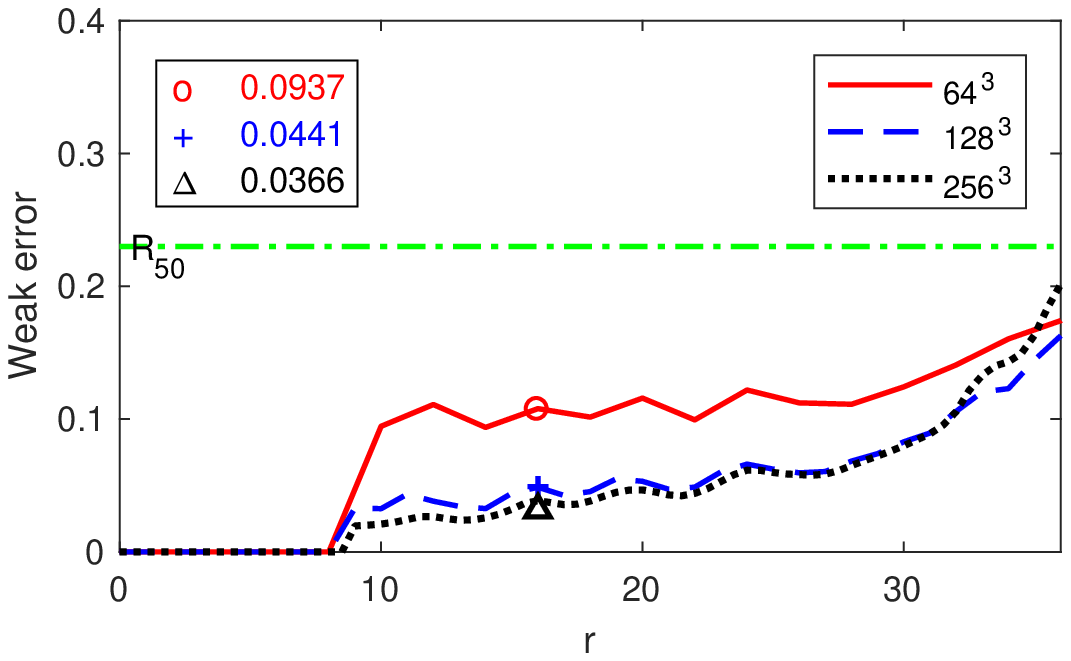}}\\
  \caption{\emph{Top left}: $R_N$,
    \emph{top right}:  $R_T$,
    \emph{bottom left}:  $R_N \oplus R_T$, and  
    \emph{bottom right}: $R_S \oplus R_N \oplus R_T$. 
    The definitions of the errors are found  in Table~\ref{tab:error_cal}.}
\label{fig:Rt_Rn_only}
\end{figure}

Figure~\ref{fig:Rt_Rn_only} illustrates the noise error. As expected,
the noise error $R_N$ is small and flat, since the compression step
significantly reduces the Poissonian noise by taking the average. We
also observe that the noise error $R_N$ is positively correlated to
the grid sizes $\Mgrid$ and the shell distance $r$, since the overall
signal contribution per voxel decreases with $r$.

A similar analysis holds for  $R_T$ and for $R_N \oplus R_T$, which are also 
positively correlated to the grid sizes $\Mgrid$ and the shell distance $r$, as 
shown in Figure~\ref{fig:Rt_Rn_only}. The rotational error $R_T$ increases with 
increasing shell distance $r$, since the error in estimating a rotational 
probability induces a larger contribution to total errors for voxels that are 
further away from the origin.
 
The last figure of Figure~\ref{fig:Rt_Rn_only} shows the combination of the 
smearing, the noise, and the rotational error. As expected, this combinated 
error again increases with increasing shell distance $r$. However, it is 
\emph{negatively} correlated to the grid sizes, since the smearing error $R_S$ 
reduces much quicker than the other errors increases. The $64^3$ resolution 
fails to perform well, due to the dominating smearing error.

Again, due to the artifacts of the truncated domain, all the errors 
investigated above increase dramatically at $r>30$.

\subsubsection{The fluence error $R_F$ and its combinations with other errors.}

\begin{figure}[!htb]
  \subfloat{\includegraphics[width=.5\textwidth]{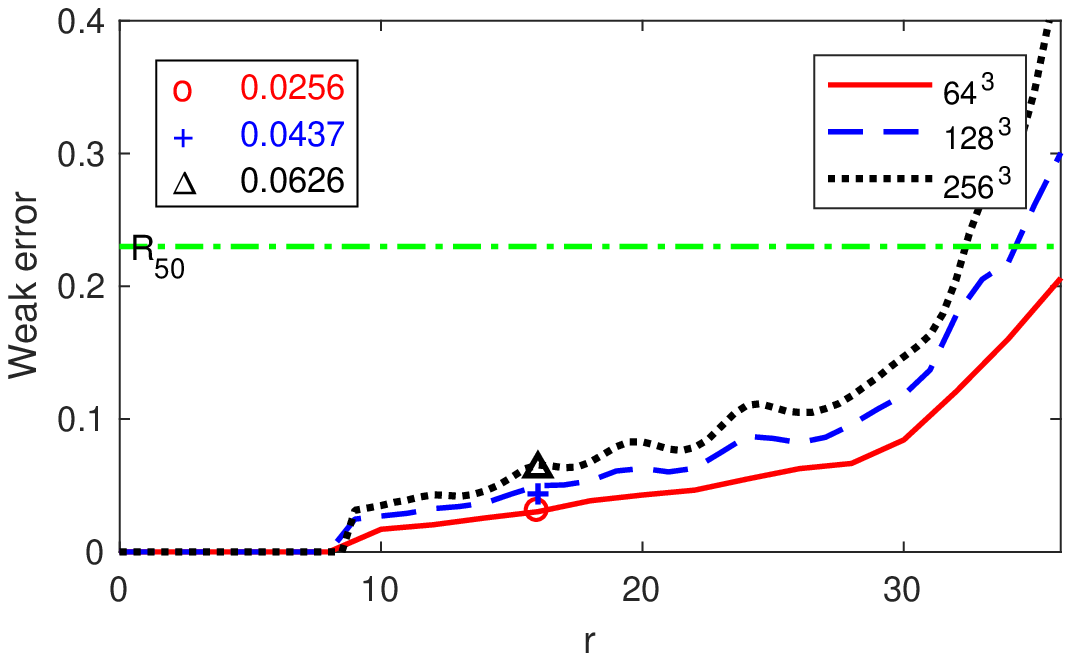}}
  \subfloat{\includegraphics[width=.5\textwidth]{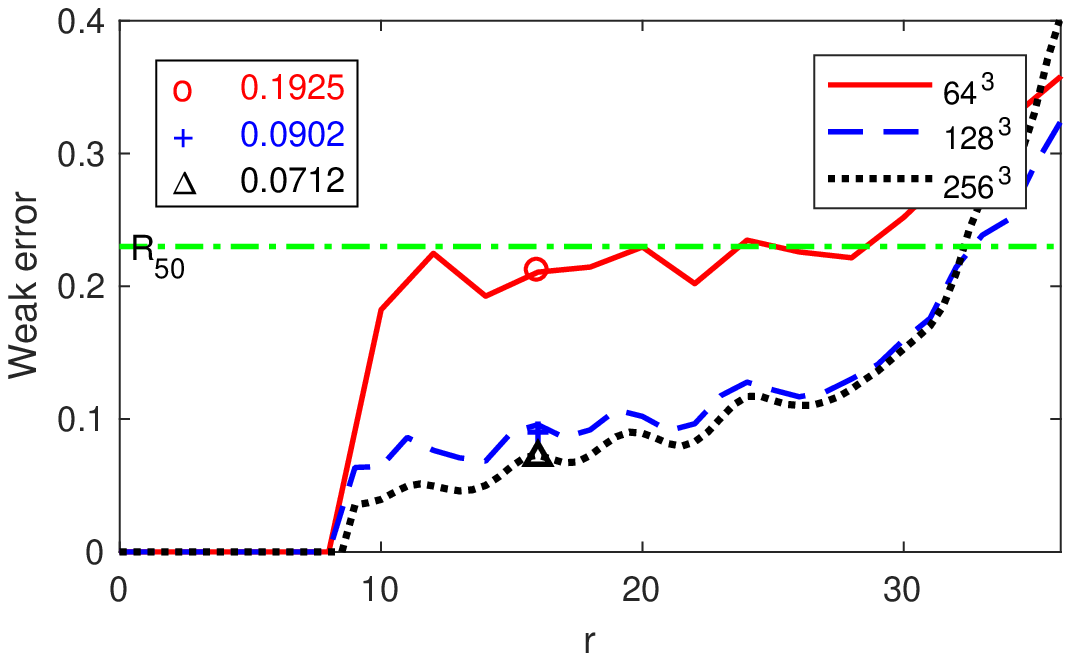}}\\
  \caption{The fluence error combinations: $R_F \oplus R_N \oplus R_T$
    (\emph{left}) and $R_S \oplus R_F \oplus R_N \oplus R_T$
    (\emph{right}). }
  \label{fig:Rfs}
\end{figure}

Finally, we measured the error induced by the fluence estimation,
i.e.~the fluence error $R_F$, and its combinations with the other
algorithmic errors. In isolation, the fluence error $R_F$ behaves
similarly to the noise error $R_N$ (not shown). Figure~\ref{fig:Rfs}
shows composite errors including $R_F$. Similar to $R_N \oplus R_T$,
the combined error $R_F \oplus R_N \oplus R_T$ correlates positively
with the grid size $\Mgrid$ and the shell distance $r$, however it is
slightly larger than $R_N \oplus R_T$. Once the smearing error $R_S$
is considered, the $R_S$ error again dominates at the
$64^3$-resolution. Further, the average of $R_S \oplus R_F \oplus R_N
\oplus R_T$ at $\Mgrid =64^3$ and $r\in(8,30)$ is 0.22, which is just
below the average ${R}_{50}$.

\subsection{Sharpness of bootstrapping}
\label{subsec:sharpness}

In this section, we estimate the reconstruction uncertainties when the correct 
information: the 3D `truth' $\Wgrid^{\perp}$, the correct fluence $\phi^*$, and 
the correct rotational probability $P^*$ are \emph{not} accessible. We 
discussed both the standard bootstrap and the EMB method in detail in \S
\ref{subsec:boot_theory}, where the total reconstruction uncertainty $R_{total}
$ was defined in \eqref{eq:boot_val}. To put $R_{total}$ in a similar form as 
the weak error metric \eqref{eq:weak_error}, we transfer $R_{total}$ to the 
following radial-shell error metric:
\begin{align}
  \tilde{e}_{u}(R_{total}, \Wgrid_a) &= \frac{\sum_{s\in S_{u}}
     {|R_{total}|_s}}{\sum_{s \in S_{u}} |\Wgrid_a|_s},
      \label{eq:boot_weak_error}
\end{align}
where $\Wgrid_a$ is reconstructed from the bootstrap universe.

To validate our bootstrap estimators, we used the fluence-affected
Poissonian signal $K^{f0}$ as the bootstrap universe. For both
bootstrap schemes, $B=100$ bootstrap samples were used, and each
sample contained $\Mdata = 1000$ frames.

\begin{figure}[!htb]
  \subfloat{\includegraphics[width=.5\textwidth]{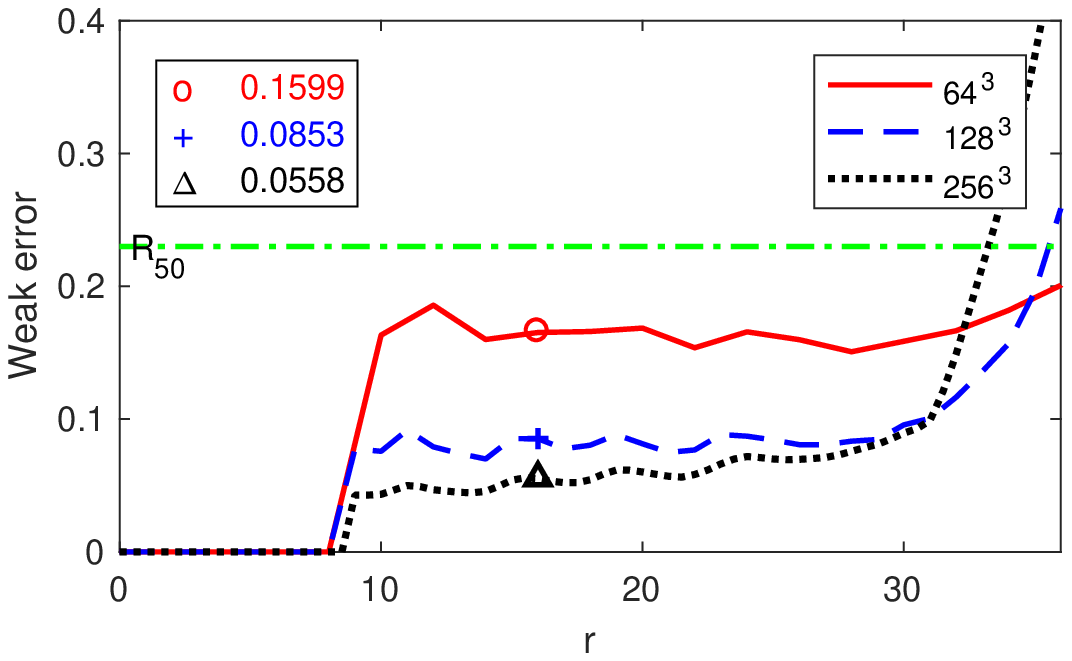}}
  \subfloat{\includegraphics[width=.5\textwidth]{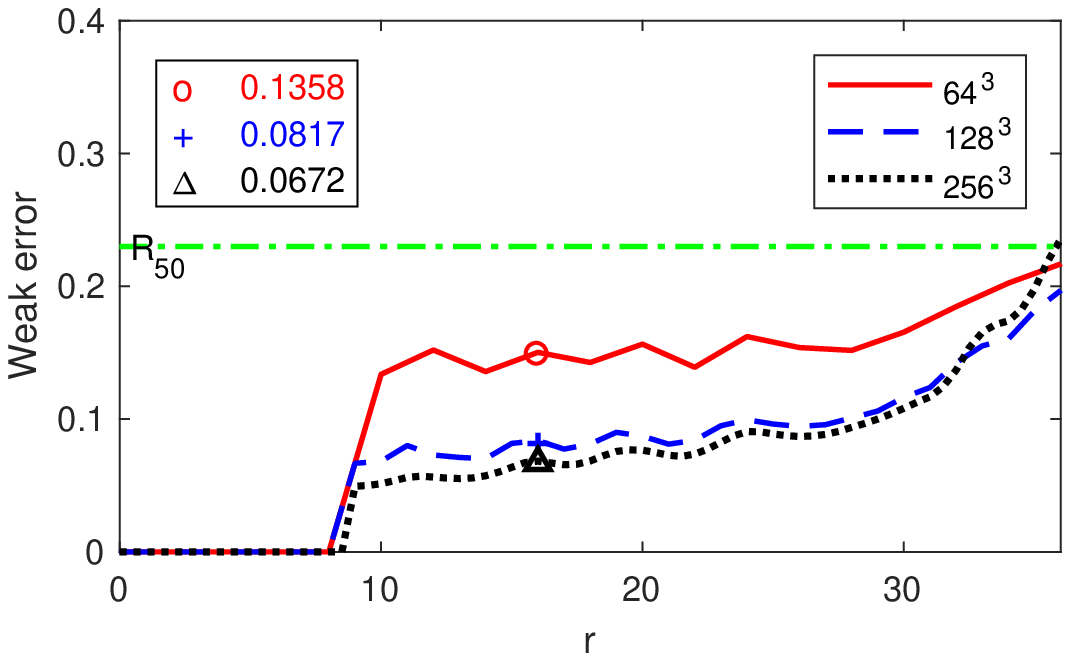}}\\
  \caption{The sharpness of the bootstrap estimators in the error
    metric \eqref{eq:boot_weak_error}. \emph{Left}: the estimated
    total reconstruction uncertainty $R_{total}$ calculated by the
    standard bootstrap estimator, and \emph{right}: $R_{total}$
    computed via the EMB estimator. Compare with the results in
    Figure~\ref{fig:Rfs}. }
  \label{fig:boot_est}
\end{figure}

Figure~\ref{fig:boot_est} shows the sharpness of the bootstrap
estimators presented in the form of the radial-shell error metric
\eqref{eq:boot_weak_error}. Comparing the results to $R_S \oplus R_F
\oplus R_N \oplus R_T$ in Figure~\ref{fig:Rfs}, both the standard
bootstrap and the EMB method produce accurate estimations at $\Mgrid =
[128^3, 256^3]$. However, at $\Mgrid = 64^3$, both estimations of
$R_{total}$ are smaller than $R_S \oplus R_F \oplus R_N \oplus
R_T$. This is due to the underestimation of the smearing error,
i.e.~$R_S$ being much larger than the estimated smearing error $\hat
R_S$~\eqref{eq:stdboot_rs}.

\subsection{Robustness for background noise} 
\label{subsec:robustness}

Other than the shot noise, the captured diffraction patterns of a
typical FXI experiment might also contain artifacts of background
noise, detector saturation, erroneous pixels, etc. In this section, we
investigate the influence of background noise with different pattern
intensities. Since the diffraction patterns are collected from the
same experimental setup, it is reasonable to assume that the background 
noise at each pixel is approximately constant from shot to shot. Hence 
we sample our data as follows,
\begin{align}
K_m \sim \po(c  \phi  K^*   + t K_{bg}),
\end{align}
where $K_{bg}$ is the background signal, which was measured from a
real FXI single-particle experiment. If $t=1$, the so generated
patterns contain added background noise. Again, $K^*$ stands for the
noiseless patterns, and $\phi \in(0.9,1.2)$. Further, $c$ is the
intensity factor that controls the total number of photons of the
diffraction pattern, and set by us in order to perform the
experiments.

To investigate the influences of intensity and the background noise,
we generated 6 datasets without background noise and with $1000$
frames in each dataset. The intensity factors $c$ of these datasets
were chosen such that the maximum number of photons in one pixel was
$P_c = [1000, 500, 100, 90, 75, 50]$ photons. We also generated their
corresponding diffraction patterns with background noise. As a
comparison, we generated another 6+6 datasets with the same $P_c$ by
enlarging the number of frames to $5000$.

\begin{figure}[!htb]
 \subfloat{\includegraphics[width=.5\textwidth]{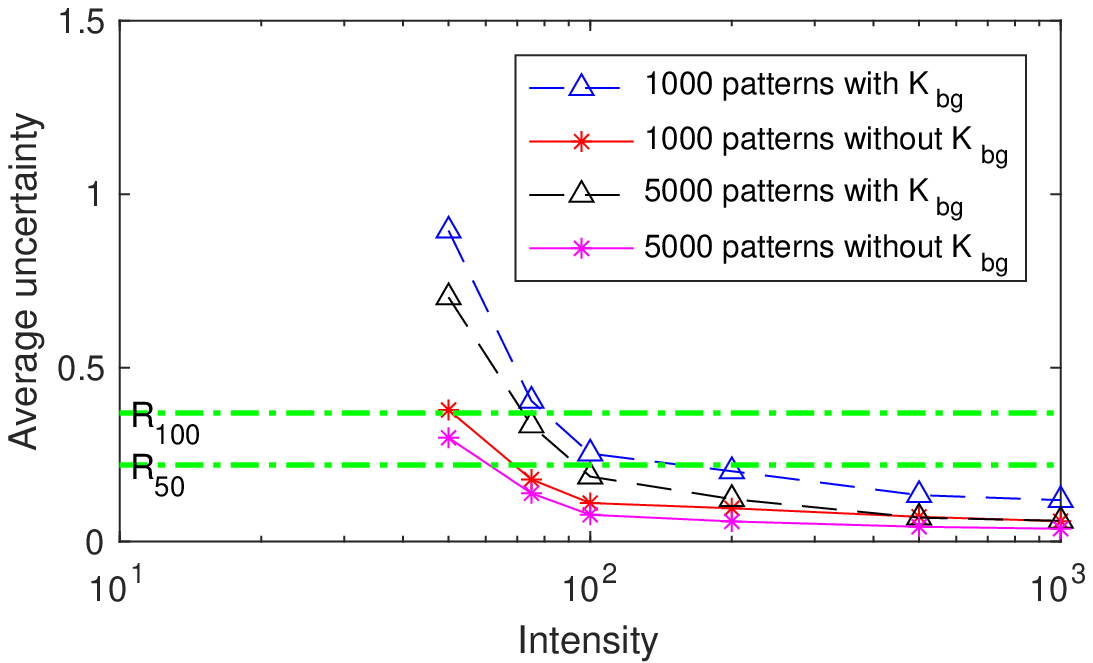}}
 \subfloat{\includegraphics[width=.5\textwidth]{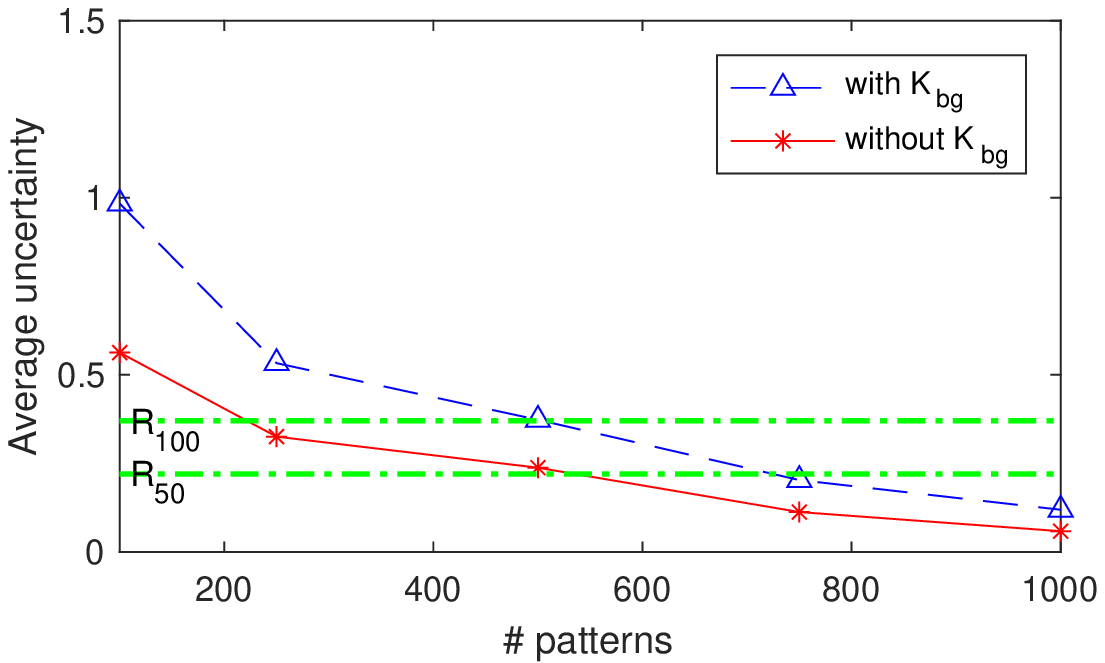}}
 \caption{\emph{Left:} The relationship between the average
   reconstruction uncertainty and the diffraction pattern intensity.
   \emph{Right:} The relationship between average reconstruction
   uncertainty and the number of frames.
 }
\label{fig:intens_loss}
\end{figure}

The standard bootstrap scheme was used to estimate the reconstruction
uncertainty $R_{total}$ for each dataset via the uncertainty estimator
\eqref{eq:boot_val}. Again $B=100$ bootstrap samples were drawn from
each dataset, and each bootstrap sample contained $\Mdata = 1000$ or
$5000$ diffraction patterns. Figure~\ref{fig:intens_loss}
(\emph{left}) shows the relationship between the intensity and the
average total reconstruction uncertainty for $\Mgrid=128^3$.

As can be seen, the background noise creates a larger reconstruction
uncertainty, since the patterns with background noise violate the
assumption of maximum likelihood \eqref{eq:ML-estimator2}.  We also
observe that the uncertainty increased with decreasing $P_c$,
especially when $P_c<100$ photons. This is due to the EMC algorithm
being unable to distinguish between the diffraction signals and the
noise. Further, increasing the number of frames for a reconstruction
reduces the total uncertainty, too.

Take a closer look at $P_c < 100$ photons in Figure
\ref{fig:intens_loss}.  The average uncertainty from 1000 diffraction
patterns is larger than the average $R_{50}$ in
Figure~\ref{fig:R_rand}. We may understand this phenomenon as being
roughly equivalent to a less than 50\% of the hidden information being
recovered from the 1000 diffraction patterns when $P_c$ is less than
100 photons. On the other hand, increasing the number of frames to
5000 reduces the uncertainty, and hence slightly more than 50\% of the
hidden information is recovered at $P_c = 100$ photons. When $P_c$ is
50 photons, no reconstruction recovers more than 50\% of the hidden
information.

We now investigate the influence of the number of frames when the
pattern signal is approximately similar to the diffraction patterns
used in the 3D reconstruction of the Mimivirus \citep{3d_mimi}, that
is when $P_c = 1000$. As can be seen from
Figure~\ref{fig:intens_loss}, the average total uncertainty reduces
with increasing number of frames. In order to obtain a reconstruction
whose uncertainty is less than $R_{100}$, we need at least 250
diffraction patterns without background noise, or 500 frames with
background noise. Further, roughly 50\% of the hidden information can
be obtained from 500 frames without background noise, or 750 frames
with background noise.

\section{Conclusions}
\label{sec:conclusions}

The FXI technique holds the promise of obtainining biological particle
structures in a near-native state without crystallization. For the
technique to become competitive with existing imaging modalities,
experiment workflow as well as algorithmic developments are
needed. Our aim has been to investigate the uncertainties of the
reconstruction procedure.

To understand the uncertainty propagation in the EMC reconstruction
procedure, we have identified several uncertainty sources and
quantitatively measured the algorithmic uncertainties with the setup
on synthetic data. For a 3D reconstruction coarsely resolved in the
diffraction space, where fringes are close to each other, the
uncertainty is high due to aliasing effects. On the other hand, the
uncertainty of a more finely resolved 3D reconstruction is low, while
time usage for the EMC algorithm will be high. The number of patterns
required for sampling the highly resolved space is also higher. Since
the uncertainty induced from the most time-consuming step of the EMC
algorithm (the rotational error) is negatively correlated to the 3D
reconstruction size, one can use the binned diffraction patterns to
calculate the rotational probability, and use the unbinned, or the
less binned patterns in the maximization step for improving the 3D
reconstruction quality as well as reducing the computation time.

In order to be relevant for more realistic cases, where the biological
particle structures are unknown, we have applied a bootstrap technique
to the reconstruction procedure for assessing the reconstruction
uncertainty. We claim that both the standard bootstrap and the EMB
estimator proposed by us work well. Furthermore, in our experiments
both bootstrap procedures are robust, and can tolerate the presence of
non-Poissonian noise. However, we recommend to use the standard
bootstrap method if the statistical model does not fit the diffraction
patterns. If the diffraction patterns are extremely noisy, it is
possible to modify the statistical model underlying the
maximum-likelihood estimate in the M step by using a penalty function,
or by directly modifying the probability distribution to account for
the presence of noise photons \cite{pan1,pan2}.

Although X-FEL science has progressed from vision to reality, and
imaging techniques are improving, high-resolution 3D structures of
single particles are still absent. For existing datasets, our findings
indicate the benefits of using a higher number of diffraction
patterns, avoiding too radical downsampling. The sampling level
appropriate for proper use of EMC is higher than the Nyquist criterion
on the level of oversampling necessary for 3D phase retrieval. The
latter criterion is the one primarily used in previous literature
discussing attainable resolution. By properly pre-processing existing
datasets, more patterns usable for 3D reconstruction can probably be
identified in many cases. Another option for attaining the sampling
necessary would be to apply symmetry or blurring/smoothing in the
compression step.

Newer facilities, such as the European XFEL, aim to increase the data
rates.  The European XFEL will be capable of acquiring 27,000
high-quality diffraction patterns per second - 225 times faster than
the Linac Coherent Light Source (LCLS) and more than 450 times faster
than the Spring-8 $\angstrom$ Compact free electron LAser
(SACLA). Through an improved understanding of the uncertainty
propagation properties of EMC, we hope that these future facilities
will, in time, allow the 3D reconstruction of individual reproducible
biological particles down to sub-nanometer resolution, with
appropriate estimates of the uncertainty in those reconstructions.

	
\section*{Acknowledgment}
	
This work was financially supported by by the Swedish Research Council
within the UPMARC Linnaeus center of Excellence (S.~Engblom, J.~Liu)
and by the Swedish Research Council, the R\"ontgen-\AA ngstr\" om
Cluster, the Knut och Alice Wallenbergs Stiftelse, the European
Research Council (J.~Liu).
	
	
\newcommand{\doi}[1]{\href{http://dx.doi.org/#1}{doi:#1}}

\newcommand{\available}[1]{Available at \url{#1}}
\newcommand{\availablet}[2]{Available at \href{#1}{#2}}

\bibliographystyle{abbrvnat} 
\bibliography{eel}	

\begin{thebibliography}{31}
\providecommand{\natexlab}[1]{#1}
\providecommand{\url}[1]{\texttt{#1}}
\expandafter\ifx\csname urlstyle\endcsname\relax
  \providecommand{\doi}[1]{doi: #1}\else
  \providecommand{\doi}{doi: \begingroup \urlstyle{rm}\Url}\fi

\bibitem[Antonelli~M. et~al.(2016)Antonelli~M., G., et~al.]{detector}
G.~T. Antonelli~M., B.~G., et~al.
\newblock Fast broad-band photon detector based on quantum well devices and
  charge-integrating electronics for non-invasive fel monitoring.
\newblock \emph{AIP Conference Proceedings}, 1741, 2016.
\newblock \doi{10.1063/1.4952824}.

\bibitem[Bishop et~al.(1998)Bishop, Svens\'en, and Williams]{gtm}
C.~M. Bishop, M.~Svens\'en, and C.~K.~I. Williams.
\newblock {GTM}: The generative topographic mapping.
\newblock \emph{Neural Computation}, 10\penalty0 (1):\penalty0 215--234, 1998.
\newblock \doi{10.1162/089976698300017953}.

\bibitem[Bogan et~al.(2008)Bogan, Benner, Boutet, et~al.]{44491}
J.~M. Bogan, W.~H. Benner, S.~Boutet, et~al.
\newblock Single particle {X-ray} diffractive imaging.
\newblock \emph{Nano letters}, 8\penalty0 (1):\penalty0 310--316, 2008.
\newblock \doi{10.1021/nl072728k}.

\bibitem[Bozek(2009)]{Bozek2009}
J.~D. Bozek.
\newblock {AMO instrumentation for the LCLS X-ray FEL}.
\newblock \emph{{The European Physical Journal Special Topics}}, 169\penalty0
  (1):\penalty0 129--132, mar 2009.
\newblock ISSN 1951-6355.
\newblock \doi {10.1140/epjst/e2009-00982-y}.

\bibitem[Chalupsk\'{y} et~al.(2007)Chalupsk\'{y}, Juha, Kuba, et~al.]{394011}
J.~Chalupsk\'{y}, L.~Juha, J.~Kuba, et~al.
\newblock Characteristics of focused soft {X-ray} free-electron laser beam
  determined by ablation of organic molecular solids.
\newblock \emph{Optics Express}, 15\penalty0 (10):\penalty0 6036--6043, 2007.
\newblock \doi{10.1364/OE.15.006036}.

\bibitem[Coifman and Lafon(2006)]{dm}
R.~R. Coifman and S.~Lafon.
\newblock Diffusion maps.
\newblock \emph{Applied and computational harmonic analysis}, 21\penalty0
  (1):\penalty0 5--30, 2006.
\newblock \doi{10.1016/j.acha.2006.04.006}.

\bibitem[Dikmen and Fevotte(2012)]{Dikmen2012}
O.~Dikmen and C.~Fevotte.
\newblock Maximum marginal likelihood estimation for nonnegative dictionary
  learning in the gamma-poisson model.
\newblock \emph{IEEE Transactions on Signal Processing}, 60\penalty0
  (10):\penalty0 5163--5175, 2012.
\newblock \doi{10.1109/TSP.2012.2207117}.

\bibitem[Efron(1979)]{bootstrapJack}
B.~Efron.
\newblock Bootstrap methods: Another look at the jackknife.
\newblock \emph{The Annals of Statistics}, 7\penalty0 (1):\penalty0 1--26,
  1979.
\newblock \doi{10.1214/aos/1176344552}.

\bibitem[Efron and Tibshirani(1994)]{bootstrapIntro}
B.~Efron and R.~J. Tibshirani.
\newblock \emph{An introduction to the bootstrap}.
\newblock CRC press, 1994.

\bibitem[Ekeberg et~al.(2015{\natexlab{a}})Ekeberg, Engblom, and Liu]{hpcEMC}
T.~Ekeberg, S.~Engblom, and J.~Liu.
\newblock Machine learning for ultrafast {X-ray} diffraction patterns on
  large-scale gpu clusters.
\newblock \emph{International Journal of High Performance Computing
  Applications}, pages 233--243, 2015{\natexlab{a}}.
\newblock \doi{10.1177/1094342015572030}.

\bibitem[Ekeberg et~al.(2015{\natexlab{b}})Ekeberg, Svenda, Abergel,
  et~al.]{3d_mimi}
T.~Ekeberg, M.~Svenda, C.~Abergel, et~al.
\newblock Three-dimensional reconstruction of the giant mimivirus particle with
  an {X-ray} free-electron laser.
\newblock \emph{Physical review letters}, 114\penalty0 (9), 2015{\natexlab{b}}.
\newblock \doi{10.1103/PhysRevLett.114.098102}.

\bibitem[Fessler and Rogers(1996)]{pan2}
J.~Fessler and W.~Rogers.
\newblock {Spatial resolution properties of penalized-likelihood image
  reconstruction: space-invariant tomographs}.
\newblock \emph{IEEE Transactions on Image Processing}, 5\penalty0
  (9):\penalty0 1346--1358, 1996.
\newblock ISSN 10577149.
\newblock \doi{10.1109/83.535846}.

\bibitem[Goodman(2005)]{f1}
J.~W. Goodman.
\newblock \emph{Introduction to Fourier Optics}, chapter~4, pages 63--96.
\newblock Roberts \& Company Publishers, 3 edition, 2005.

\bibitem[Hau-Riege et~al.(2007)Hau-Riege, London, Chapman, et~al.]{44499}
S.~P. Hau-Riege, R.~A. London, H.~N. Chapman, et~al.
\newblock Encapsulation and diffraction-pattern-correction methods to reduce
  the effect of damage in {X-ray} diffraction imaging of single biological
  molecules.
\newblock \emph{Physical Review Letters}, 98\penalty0 (19):\penalty0 198302,
  2007.
\newblock \doi{10.1103/PhysRevLett.98.198302}.

\bibitem[J~D~Bozek et~al.(2015)J~D~Bozek, Hui, et~al.]{amoIn}
L.~F. J~D~Bozek, J C~Castagna, Z.~Hui, et~al.
\newblock X-ray split and delay device for ultrafast {X}-ray science at the
  {AMO} instrument at {LCLS}.
\newblock \emph{Journal of Physics: Conference Series}, 635\penalty0
  (1):\penalty0 12--18, 2015.
\newblock \doi{10.1088/1742-6596/635/1/012018}.

\bibitem[Kassemeyer et~al.(2012)Kassemeyer, Steinbrener, Lomb, et~al.]{515159}
S.~Kassemeyer, J.~Steinbrener, L.~Lomb, et~al.
\newblock Femtosecond free-electron laser {X-ray} diffraction data sets for
  algorithm development.
\newblock \emph{Optics Express}, 20\penalty0 (4):\penalty0 4149--4158, 2012.
\newblock \doi{10.1364/OE.20.004149}.

\bibitem[Kirian R.~A. et~al.(2015)]{injector}
E.~N. Kirian R.~A., Awel~S. et~al.
\newblock Simple convergent-nozzle aerosol injector for single-particle
  diffractive imaging with {X}-ray free-electron lasers.
\newblock \emph{Structural Dynamics}, 2\penalty0 (4), 2015.
\newblock \doi{10.1063/1.4922648}.

\bibitem[Lee and Seung(1999)]{Lee1999}
D.~D. Lee and H.~S. Seung.
\newblock Learning the parts of objects by non-negative matrix factorization.
\newblock \emph{Nature}, 401\penalty0 (6755):\penalty0 788--791, 1999.
\newblock \doi{10.1038/44565}.

\bibitem[Lee and Seung(2001)]{Lee2001}
D.~D. Lee and H.~S. Seung.
\newblock Algorithms for non-negative matrix factorization.
\newblock In \emph{Advances in Neural Information Processing Systems}, pages
  556--562. MIT Press, 2001.

\bibitem[Loh and Elser(2009)]{EMC}
N.~D. Loh and V.~Elser.
\newblock Reconstruction algorithm for single-particle diffraction imaging
  experiments.
\newblock \emph{Physical Review E}, 80\penalty0 (2):\penalty0 026705, 2009.
\newblock \doi{10.1103/PhysRevE.80.026705}.

\bibitem[Loh et~al.(2010)Loh, Bogan, Elser, et~al.]{EMC2}
N.~D. Loh, M.~J. Bogan, V.~Elser, et~al.
\newblock Cryptotomography: Reconstructing 3{D} {F}ourier intensities from
  randomly oriented single-shot diffraction patterns.
\newblock \emph{Physical Review Letters}, 104:\penalty0 225501, 2010.
\newblock \doi{10.1103/PhysRevLett.104.225501}.

\bibitem[Maia et~al.(2010)Maia, Ekeberg, van~der Spoel, et~al.]{306934}
F.~R.~N.~C. Maia, T.~Ekeberg, D.~van~der Spoel, et~al.
\newblock Hawk : the image reconstruction package for coherent {X-ray}
  diffractive imaging.
\newblock \emph{Journal of Applied Crystallography}, 43:\penalty0 1535--1539,
  2010.
\newblock \doi{10.1107/S0021889810036083}.

\bibitem[Neutze et~al.(2000)Neutze, Wouts, van~der Spoel, et~al.]{dad}
R.~Neutze, R.~Wouts, D.~van~der Spoel, et~al.
\newblock Potential for biomolecular imaging with femtosecond {X-ray} pulses.
\newblock \emph{Nature}, 406\penalty0 (6797):\penalty0 752--757, 2000.
\newblock \doi{10.1038/35021099}.

\bibitem[(PDB)(2015)]{pdbsta}
R.~P. D.~B. (PDB).
\newblock Pdb current holdings breakdown, 6 2015.
\newblock URL \url{http://www.rcsb.org/pdb/statistics/holdings.do}.

\bibitem[Seibert et~al.(2011)Seibert, Ekeberg, Maia, et~al.]{mimivirus_Xrays}
M.~M. Seibert, T.~Ekeberg, F.~R.~N.~C. Maia, et~al.
\newblock Single mimivirus particles intercepted and imaged with an {X}-ray
  laser.
\newblock \emph{Nature}, 470\penalty0 (7332):\penalty0 78--81, 2011.
\newblock \doi{10.1038/nature09748}.

\bibitem[Ugray~Zsolt et~al.(2007)Ugray~Zsolt, John, et~al.]{optGlobal}
L.~L. Ugray~Zsolt, P.~John, et~al.
\newblock {Scatter Search and Local NLP Solvers: A Multistart Framework for
  Global Optimization}.
\newblock \emph{{INFORMS J. on Computing}}, 19\penalty0 (3):\penalty0 328--340,
  2007.
\newblock \doi{10.1287/ijoc.1060.0175}.

\bibitem[Wu(2006)]{emb1}
X.~Wu.
\newblock Incorporating large unlabeled data to enhance {EM} classification.
\newblock \emph{Journal of Intelligent Information Systems}, 26\penalty0
  (3):\penalty0 211--226, 2006.
\newblock \doi{10.1007/s10844-006-0865-3}.

\bibitem[Xu et~al.(2010)Xu, Taguchi, and W.Tsui]{pan1}
J.~Xu, K.~Taguchi, and B.~M. W.Tsui.
\newblock {Statistical Projection Completion in X-ray CT Using Consistency
  Conditions}.
\newblock \emph{IEEE Transactions on Medical Imaging}, 29\penalty0
  (8):\penalty0 1528--1540, aug 2010.
\newblock \doi{10.1109/TMI.2010.2048335}.

\bibitem[Yanez and Bach(2014)]{Yanez2014}
F.~Yanez and F.~R. Bach.
\newblock Primal-dual algorithms for non-negative matrix factorization with the
  {K}ullback-{L}eibler divergence.
\newblock \emph{CoRR}, abs/1412.1788, 2014.

\bibitem[Yang et~al.(2011)Yang, Zhang, Yuan, et~al.]{Zhi2011}
Z.~Yang, H.~Zhang, Z.~Yuan, et~al.
\newblock {K}ullback-{L}eibler divergence for nonnegative matrix factorization.
\newblock \emph{Artificial Neural Networks and Machine Learning}, pages
  250--257, 2011.
\newblock \doi{10.1007/978-3-642-21735-7\_31}.

\bibitem[Zribi(2010)]{emb2}
M.~Zribi.
\newblock Non-parametric and region-based image fusion with bootstrap sampling.
\newblock \emph{Information Fusion}, 11\penalty0 (2):\penalty0 85--94, 2010.
\newblock \doi{10.1016/j.inffus.2008.08.004}.

\end{thebibliography}

\end{document}